\documentclass[fleqn,usenatbib]{mnras}

\usepackage{newtxtext,newtxmath}
\usepackage{mathptmx}

\usepackage[T1]{fontenc}
\usepackage{ae,aecompl}

\usepackage{graphicx}	
\usepackage{amsmath}	
\usepackage{amssymb}	

\usepackage{subfigure}
\usepackage{cancel}
\usepackage[title]{appendix}

\title[Tidal disruptions of planetary bodies II]{Tidal disruption of planetary bodies by white dwarfs II: Debris disc structure and ejected interstellar asteroids}

\author[Malamud \& Perets]
{		
	Uri Malamud$^{1,2}$ and
	Hagai B. Perets$^{1,3}$
	\\	
	$^{1}$Department of Physics, Technion - Israel Institute of Technology, Technion City, 3200003 Haifa, Israel\\
	$^{2}$School of the Environment and Earth Sciences, Tel Aviv University, Ramat Aviv, 6997801 Tel Aviv, Israel\\
	$^{3}$TAPIR, California Institute of Technology, Pasadena, CA 91125, USA
}

\date{Accepted XXX. Received YYY; in original form ZZZ}

\pubyear{2019}

\begin{document}

\label{firstpage}
\pagerange{\pageref{firstpage}\textendash{}\pageref{lastpage}}
\maketitle
	
\begin{abstract}
We make use of a new hybrid method to simulate the long-term, multiple-orbit disc formation through tidal disruptions of rocky bodies by white dwarfs, at high-resolution and realistic semi-major axis. We perform the largest-yet suite of simulations for dwarf and terrestrial planets, spanning four orders of magnitude in mass, various pericentre distances and semi-major axes between 3 AU and 150 AU. This large phase space of tidal disruption conditions has not been accessible through the use of previous codes. We analyse the statistical and structural properties of the emerging debris discs, as well as the ejected unbound debris contributing to the population of interstellar asteroids. Unlike previous tidal disruption studies of small asteroids which form ring-like structures on the original orbit, we find that the tidal disruption of larger bodies usually forms dispersed structures of interlaced elliptic eccentric annuli on tighter orbits. We characterize the (typically power-law) size-distribution of the ejected interstellar bodies as well as their composition, rotation velocities and ejection velocities. We find them to be sensitive to the depth (impact parameter) of the tidal disruption. Finally, we briefly discuss possible implications of our results in explaining the peculiar variability of Tabby's star, the origin of the transit events of ZTF J0139+5245 and the formation of a planetary core around SDSS J1228+1040.

\end{abstract}
	

\section{Introduction}\label{S:Intro}
Between 25\% to 50\% of all WDs \citep{ZuckermanEtAl-2003,ZuckermanEtAl-2010,KoesterEtAl-2014} are found to be polluted with heavy elements. The existence of orbiting gas and dust as deduced from measurements of infrared excess and metal emission lines are thought to originate from accretion of planetary material \citep{DebesSigurdsson-2002,Jura-2003,KilicEtAl-2006,Jura-2008}. In particular, the inferred composition of both WD atmospheres \citep{WolffEtAl-2002,DufourEtAl-2007,DesharnaisEtAl-2008,KleinEtAl-2010,GansickeEtAl-2012,JuraYoung-2014,HarrisonEtAl-2018,HollandsEtAl-2018,DoyleEtAl-2019, SwanEtAl-2019} and their discs \citep{ReachEtAl-2005,JuraEtAl-2007,ReachEtAl-2009,JuraEtAl-2009,BergforsEtAl-2014,Farihi-2016,ManserEtAl-2016,DennihyEtAl-2018} suggests that the polluting material is terrestrial-like and typically dry.

Such material was suggested to originate from planetary bodies which are perturbed by some mechanism \citep{DebesSigurdsson-2002,BonsorEtAl-2011,DebesEtAl-2012,KratterPerets-2012,PeretsKratter-2012,ShapeeThompson-2013,MichaelyPerets-2014,VerasGansicke-2015,StoneEtAl-2015,HamersPortegiesZwart-2016,Veras-2016,PayneEtAl-2016,CaiazzoHeyl-2017,PayneEtAl-2017,PetrovichMunoz-2017,StephanEtAl-2017,SmallwoodEtAl-2018} to highly eccentric orbits with proximity to the WD, and are subsequently tidally disrupted to form a circumstellar disc of planetary debris.

	
	
Neither the original processes that form debris disks around WDs  \citep{VerasEtAl-2014}, nor the subsequent processes that generate a more compact accretion disc \citep{VerasEtAl-2015} nor the actual accretion processes onto the WD \citep{Jura-2008,Rafikov-2011,MetzgerEtAl-2012} are entirely understood. The former in particular, is still very much an open question. We do not know how the original debris disc initially forms, or in what way are the debris properties affected by the orbit, eccentricity, size and composition of the tidally disrupted progenitor. We are also driven by recent discoveries of disintegrating minor planets with short \citep{VanderburgEtAl-2015} as well as much longer \citep{VanderboschEtAl-2019} periodicity, the possible leftover core of a planet whose outer layers have been removed \citep{ManserEtAl-2019}. These discoveries have generated a wide interest that extends beyond the white dwarf and planet-formation communities, and which emphasize the potential importance of understanding the initial stages of tidal disruption and debris disc formation.

To date, there exist very few detailed simulations of disc formation through tidal disruptions by WDs \citep{DebesEtAl-2012,VerasEtAl-2014}, and these are limited in resolution and length due to the computational expense of the hydrodynamical/N-body simulations involved. In an accompanying paper (Malamud \& Perets 2019; hereafter Paper I) we have presented a novel hybrid hydrodynamical-analytical approach that can resolve such difficulties. In this paper we utilize the code to perform the largest-yet suite of simulations of tidal disruptions of dwarf and terrestrial planets by WDs, covering a large set of orbital setups, and a very wide range of masses up to terrestrial planet size. This phase space probes for the first time the formation sequence, consequences and ensuing properties of debris discs that emerge in such tidal disruptions. Without the hybrid model, such initial conditions were never before accessible through the use of any previous code.

In paper I, we discuss the distinctive difference between debris discs of small versus large disrupted planetesimals. Given characteristic Solar system distances, small asteroids usually form a narrow ring in which all the asteroid material is bound to the star and remains on semi-major axes close to that of the original projenitor (in similarity to the \cite{VerasEtAl-2014} study). Such disruptions are therefore termed 'non-dispersive regime'. In contrast, larger dwarf or terrestrial sized planets usually form a completely different debris disc, in which just over half of the material becomes tightly bound to the star (compared to the original progenitor orbit) and the other half becomes unbound. Such disruptions are therefore termed 'bi-modal regime'. This paper is the first to present detailed simulations of the latter type.

The layout of the paper is structured as follows: in Section \ref{S:hybrid-approach} we briefly review our hybrid approach, and the criteria for selecting a suite of hybrid simulations of tidal disruptions by WDs, spanning a large range of masses, semi-major axes and pericentre distances. In Section \ref{S:Disc} we track the formation process of the ensuing debris discs, characterizing their typical formation timescales, and discussing their statistical properties, including fragment orbits, sizes and rotations. In Section \ref{S:Unbound} we track the ejected unbound debris that form interstellar asteroids. We likewise discuss the statistics of their size distribution, rotation and velocity. In Section \ref{S:Future} we briefly discuss possible implications and future work regarding the origin of the variability of Tabby's star, the transiting object ZTF J0139+5245, and the formation of what could be a planetary core remnant in SDSS J1228+1040, potentially arising from tidal disruptions of dwarf or terrestrial sized planets. Finally, in Section \ref{S:Summary} we summarize the paper's main points.

\section{The hybrid approach: a study of tidally disrupted dwarf and terrestrial planets}\label{S:hybrid-approach}
\subsection{Outline}\label{SS:outline}
The hybrid approach is intended for an efficient treatment of disc formation by tidal disruptions. The details of this approach are introduced and discussed in detail in Paper I. Here we only provide a brief review of the main ideas.
 
The approach makes use of the fact that the primary processes taking place during these tidal disruptions are restricted to a relatively small spatial domain. First, we recall that the differential gravitational force that breaks the object apart is relevant only to the Roche limit of the star. The second important phase is fragmentation. It is during this phase, that small particles in the tidal stream may collapse by the stream's own self-gravity, to form larger fragments. The relevant spatial domain here, as discussed in Paper I, is also rather small, exceeding the tidal sphere only by an order of magnitude or so. Overall, the breakup and fragmentation phases constitute only a tiny fraction of the total spatial domain (of the original orbit), and are confined to the immediate environment of the star.

Our approach is therefore to restrict the SPH computations only to this relatively small domain, and to omit unnecessary calculations outside of it. Following the fragmentation phase, we identify the emerging fragments (whose constituent particles form spatially connected clumps of material), and for the reminder of their orbits, their trajectories are calculated and tracked analytically, assuming Keplerian orbits. Our hybrid program simply places each fragment once again near the star's Roche limit, based on its return orbital elements. This approach computationally outperforms full SPH modeling by at least a few orders of magnitude, and the benefit increases for more eccentric orbits. We make the assumption that the disc of debris is largely collisionless (shown to be accurate at the 99\% level in Paper I), as well as dynamically unaffected by radiation or other processes. The returning fragment immediately undergoes an additional tidal disruption, potentially splitting into a new set of fragments, with their own unique dispersion in orbital parameters, and so on, in an iterative process.

The hybrid code's main task is to identify the fragments, accurately calculate their orbits and especially handle the synchronization and timing of the subsequent disruptions. Apart from this, its other procedural task is handling the SPH job dissemination. The hybrid code terminates when reaching one of two outcomes: either all fragments have ceased disrupting given their exact size, composition and orbit; or fragment disruption is inhibited when reaching its minimum size - that of a single SPH particle.

\subsection{Simulation suite and code setup}\label{SS:Suite}
The only previous work to simulate tidal disruptions and disc formation around WDs, considered small, kilometre-sized asteroids \citep{VerasEtAl-2014}. Using the same modified N-body simulation model and a similar setup, disruptions of larger objects around a WD were also investigated by the same authors \citep{VerasEtAl-2017}, however they do not discuss the formation of a disc or its emerging features, but rather focus on the rate of mass shedding of a differentiated planetesimal. Here we perform a complimentary investigation, studying for the first time objects that range in size from small dwarf-planets to terrestrial-sized planets, while focusing on the emerging debris discs. We note that the tail in the mass distribution function observed in WD atmospheres (see Figure 6 in \citep{Veras-2016}) does not currently reach beyond the mass of a typical dwarf-planet, however this could merely be an observational constraint. Certainly the large and dense orbiting object from \cite{ManserEtAl-2019} implies that more massive planet or dwarf-planet disruptions are a possibility.

For our suite of simulations, whose results are summarized in Section \ref{S:Disc} and \ref{S:Unbound} below, we consider multiple pericentre distances: 0.1$R_{\odot}$, 0.5$R_{\odot}$ and 1$R_{\odot}$. The significance of the pericentre distance is discussed in length in Paper I, where it was shown that a large pericentre distance close to the Roche limit, results in a partial disruption, shedding only a small fraction of the planet's mass, emanating from its outer portions. Whereas if the pericentre is halved, the disruption is full. It forms a classic narrow stream of debris, however the stream is gravitationally self-confined, and it collapses to form multiple fragments, which later also collide and merge among themselves, forming smaller second-generation particles. If the pericentre is merely a small fraction of the Roche limit, the disruption is both full and violent, such that the stream is gravitationally unconfined and the planet becomes almost entirely disassociated into its constituent particles. 

Here we focus primarily on partial disruptions from large pericentre distances for the following reasons. As mentioned in Section \ref{S:Intro}, numerous possibilities have been suggested regarding the mechanisms that inject planetesimals to tidal crossing orbits. At our current level of understanding however, it is not yet possible to quantify their relative importance. Hence, we cannot estimate what choice of $q$ is physically more judicious or probable than any other, arbitrary choice. Generally speaking, studies of planetesimal injection rates (through the use of loss-cone formalism, e.g. \cite{RickmanEtAl-2008,GrishinVeras-2019} and references therein) in the fast diffusion, full loss-cone regime suggest a linear dependence on the closest approach \citep{PeretsGualandris-2010}. However, in the empty loss-cone regime, planetesimals diffuse slowly and therefore typically go through Roche grazing orbits before further diffusing into more radial orbits. In other words, at least in the empty loss-cone regime, deeper disruptions are more depleted compared with the full loss-cone regime, and our choice to focus on large $q$ therefore gains more credibility. Finally, based on Paper I, we know that the hybrid model performs very well for large $q$. If one is to choose arbitrarily, it is plausible to focus on a high value of $q$, since the disc formation cannot otherwise be studied by any other means or existing models.

The fiducial WD mass in all of our simulations is 0.6 $M_{\odot}$, a common choice in many WD studies \citep{VerasEtAl-2014}. We consider the objects listed in Table \ref{tab:HybridModels}, as \emph{approximate} Solar system analogues of dry composition, noting that we are not at all concerned with simulating their precise physical properties. We merely wish to cover a large range of masses and orbits, varying by several orders of magnitude.

\begin{table}
	\caption{Hybrid suite of simulations.}
	\begin{tabular}{*{4}{l|}}	
		\hline
	 	{Solar system analogue} & {Mass ($M_{\oplus}$)} & {a (AU)} & {q ($R_{\odot}$)}\\
		\hline
		\emph{Earth}		& 1 	& 3 & 0.1,~~0.5,~~1\\
		\emph{Mars}  & 0.1 & 3 & 0.1,~~0.5,~~1\\
		\emph{Main-belt (Vesta)} & 0.0001 & 9 & 1\\
		\emph{Satellite (Io)} & 0.01 & 17 & 1\\
		\emph{Kuiper-belt (Eris)} & 0.001 & 150 & 1\\
		\hline
	\end{tabular}
     \label{tab:HybridModels}
	The listed masses are roughly compatible with our Solar system analogues. The semi-major axis $a$ is estimated assuming orbital expansion of approximately thrice the original orbit, and the pericentre distance $q$ is given in units of the WD's Roche radius.
\end{table}

For our Solar system analogues we select Earth (terrestrial-sized), Mars (planetary embryo sized), Vesta (main-belt rocky asteroid), Io (rocky satellite) and Eris (rocky Kuiper-belt object). Coincidentally, their masses vary by approximately one order of magnitude each, covering four orders of magnitude relative to the Earth's mass. 

The distances listed in Table \ref{tab:HybridModels} are the final semi-major axes during the WD stellar evolutionary phase. At $a=3$ AU, the orbit roughly corresponds to the maximum stellar radius of a $\sim$2$M_\odot$ star (a 0.6 $M_\odot$ WD progenitor) during the AGB phase \citep{MustillVillaver-2012}. A planetesimal orbiting at 1 AU will undergo orbital expansion of a factor of $\sim3$ ($2/0.6$), hence it has the potential to survive to the WD stage. We thus consider 3 AU to be the \emph{lowest possible} semi-major axis permitted for these simulations. According to more detailed analyses by \cite{MustillVillaver-2012}, Earth-like planets undergoing adiabatic orbital expansion may only survive engulfment by a 2$M_\odot$ thermally pulsating AGB star when their initial orbit is beyond about 2 AU. Hence, they are more likely to have a final semi-major axis of 6-7 AU. We further note that orbital decay by tidal interaction during the much longer RGB phase is not important for terrestrial planets, since according to previous studies they do not experience strong tidal forces as do gas giants \citep{VillaverEtAl-2014} (but the gas-giant migration can still affect the orbital stability of smaller terrestrial planets). Here we shall ignore all such complexities and simply select the smallest, physically permitted semi-major axis of 3 AU, to be used as the minimum semi-major axis throughout this paper. The subsequent semi-major axes in Table \ref{tab:HybridModels} therefore correspond to the main-sequence orbits of 1, 2.7, 5 and 45 AU, respectively.

The details concerning the \emph{miluphCUDA} SPH code setup, which is used to model the disruptions in the hybrid model, are identical to the those used in the full SPH simulations in Paper I (see details therein). For simplicity, we assume the same 'terrestrial-like' composition and structure for all the analogue objects listed in Table \ref{tab:HybridModels}. I.e, they are all composed of a rocky mantle and an iron core (with 70\% and 30\% mass fractions respectively), despite obvious differences between them that might render their composition, in reality, rather dissimilar (especially the Kuiper-belt analogue).

Finally, the resolution used in the hybrid model for this study ranges between 50K and 500K SPH particles. It is orders of magnitude higher than that used in the full SPH simulations discussed in Paper I, as indeed any other previous disc formation study. In most of the simulations below we use a nominal resolution of 200K SPH particles.

\section{Disc formation and properties}\label{S:Disc}
\subsection{Semi-major axis distribution}\label{SS:Semi-majorAxis}
\begin{figure}
	\begin{center}		
     	\subfigure[q=0.1$R_{\odot}$] {\label{fig:a0_1q3a1mI}\includegraphics[scale=0.5]{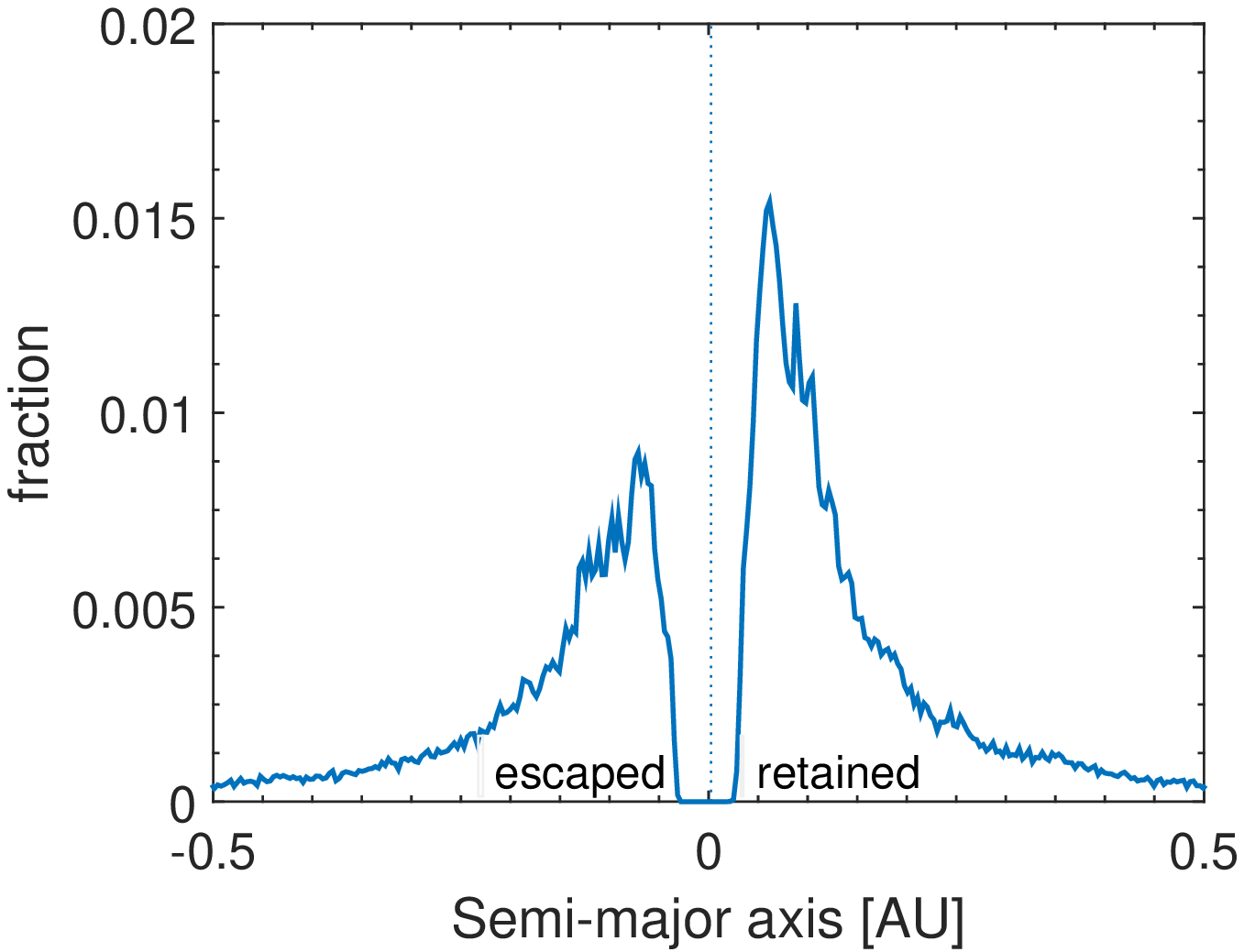}}
 		\subfigure[q=0.5$R_{\odot}$] {\label{fig:a0_5q3a1mI}\includegraphics[scale=0.5]{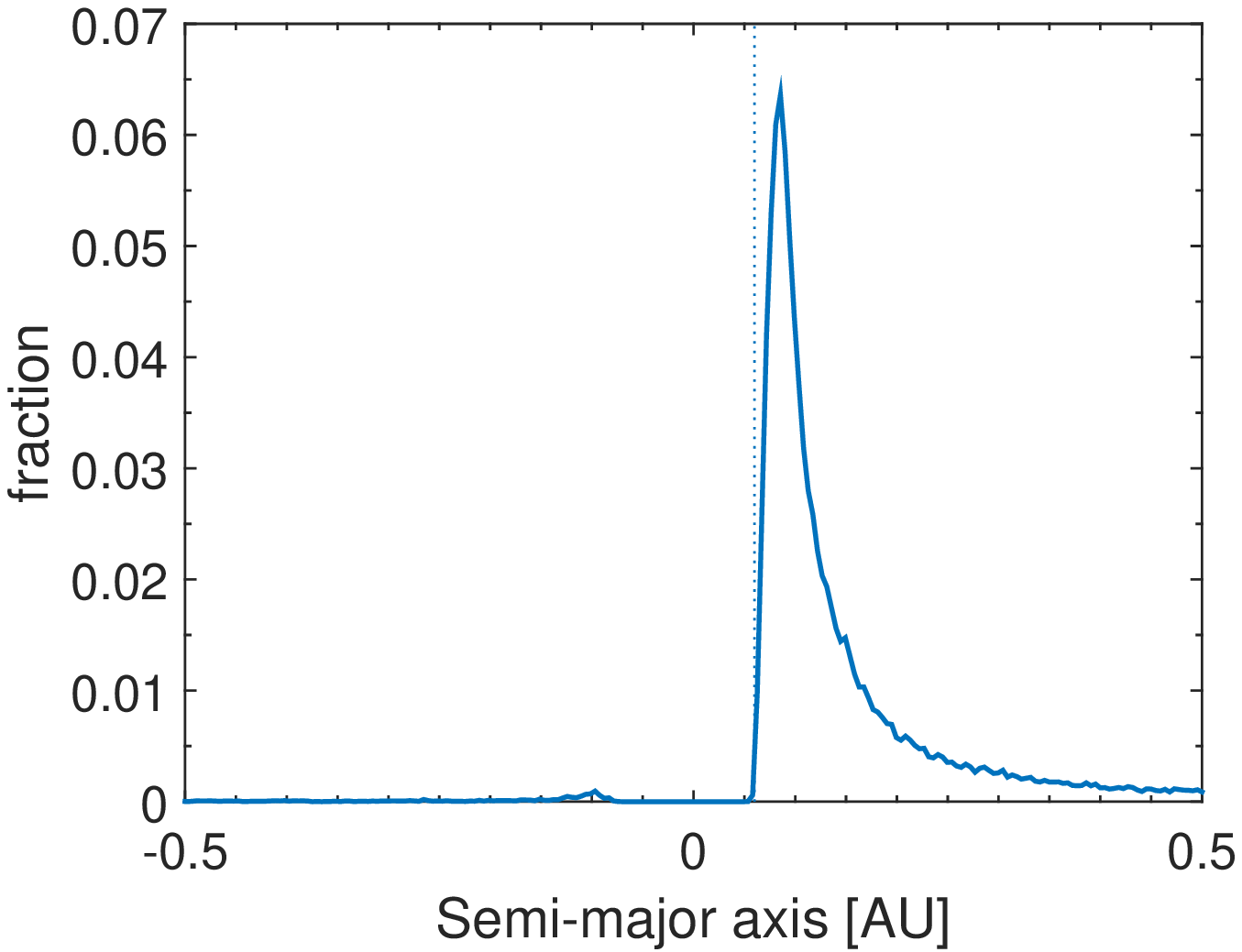}}
	 	\subfigure[q=1$R_{\odot}$]
		 {\label{fig:a1q3a1mI}\includegraphics[scale=0.5]{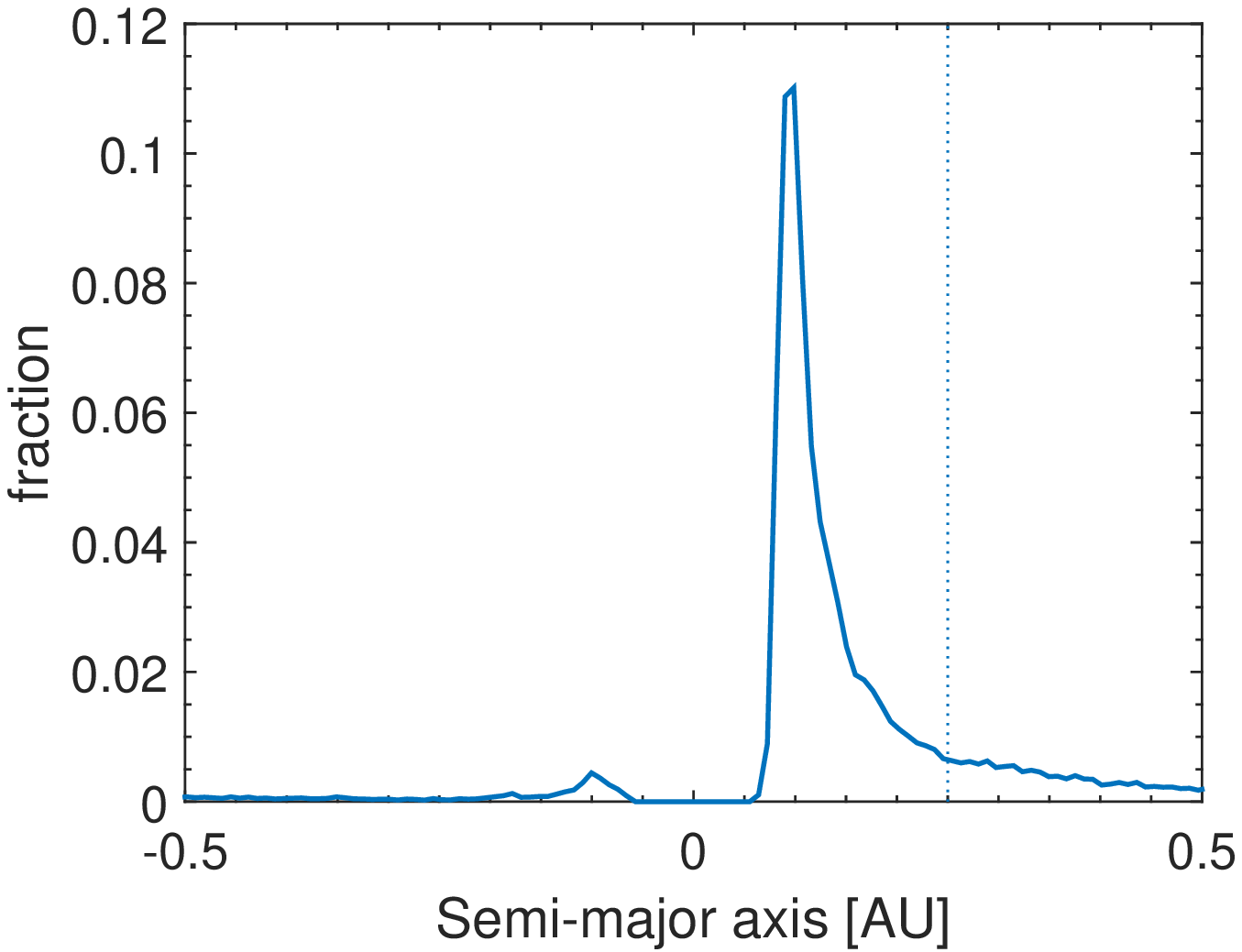}}	
	\end{center}
	\caption{Semi-major axis normalized histogram of the final debris disc, formed by the tidal disruption of an Earth-sized planet around a  0.6$M_\odot$ WD. The planet original Semi-major axis is 3 AU, whereas its pericentre distance $q$ is (a) $0.1R_{\odot}$; (b) $0.5R_{\odot}$ and (c) $1R_{\odot}$, respectively. Resolution is 200K particles. Fragments retained ($a>0$) or escaped ($a<0$) in the disruption are labelled in Panel (a). The theoretical impulse approximation model limits for the innermost semi-major axis are shown by vertical dotted lines. For a top-down view of the corresponding (bound) disc layout and temporal evolution see Paper I, Figure 8.}
	\label{fig:a3auI}
\end{figure}

We analyse the histogram distribution of the semi-major axis $a$ for various simulations. In Figure \ref{fig:a3auI}, we first look at simulations of Earth-sized planets, comparing their histograms (normalized by the total number of disrupted fragments) for different values of the pericentre distance $q$. Here the planet's original semi-major axis is 3 AU, and the WD mass is 0.6$M_\odot$. We note that these histograms capture the debris state at the \emph{end} of the simulations. Hence, they may represent one major disruption event, as in the case of Panel \ref{fig:a0_1q3a1mI}; a superposition of the planet disruption and a multitude of fragment disruptions, corresponding to the full yet gravitationally self-confined state represented in Panel \ref{fig:a0_5q3a1mI}; or a superposition of several partial disruptions that rip only of the outer portions of the original planet, and eventually the entire planet, as in Panel \ref{fig:a1q3a1mI}. Such distinctions are important for our interpretation of the histogram results, and in Section \ref{SS:SizeDistribution} we further discuss the temporal evolution.

In Paper I, we introduce an analytical impulse approximateion model, where we denote the planetesimal radius as $R$, and the displacement of its constituent particles as $r$, where $r_\mathrm{crit}$ is the critical displacement from the planetesimal's centre of mass in the opposite direction of the WD, beyond which all constituent particles will become unbound after the disruption. For an Earth-like planet, $r_\mathrm{crit}\ll R$ is a reasonable approximation. Hence, we can associate such disruptions with the the \emph{bi-modal disruption regime} (see Section \ref{S:Intro}), and calculate the tightest (innermost) semi-major axis of the disrupted constituent particles ($\acute{a}$) in the ensuing disc according to the analytical model. In the simple impulse approximation view, the innermost $\acute{a}$ arises from the radial separation of the particles on the planet's surface, and equals $d^2/2R$, where $d$ is the breakup distance. Plugging in the relevant numbers, and replacing $d$ for $q$, only the inner cut-off in Panel \ref{fig:a0_5q3a1mI} agrees with the analytical expectations (dotted vertical lines), as the innermost semi-major axis is calculated to be $\sim 0.06$ AU, just like in the plot. For the deep disruption the analytical calculation leads to $\sim 0.0025$ AU, whereas Panel \ref{fig:a0_1q3a1mI} cuts-off at a semi-major axis around 10 times larger. For the grazing disruption, the analytical model leads to $\sim 0.25$ AU, whereas Panel \ref{fig:a1q3a1mI} cuts-off at a semi-major axis of only $0.05$ AU, around 5 times smaller.

The reason for this discrepancy is not trivial. We suggest three potential causes. The impulse approximation analytical model assumes that at the moment of breakup, the body's constituent particles instantaneously disassemble from the spherical rigid body, and instead form a swarm of particles with the same velocity as prior to the breakup ($\acute{v}=v$), but now evolve independently according to their spatial distribution. In reality, even if we assume the planet to be non-rotating prior to the disruption, simulations suggest that none of these assumptions is exactly correct. (a) The precise moment of breakup is difficult to determine, but it may actually precede the moment of closest approach (hence affecting $d$); (b) The breakup is not instantaneous, and thus to some extant the object deviates from sphericity when this moment is reached (hence affecting $R$); (c) The non-instantaneous transformation from the gravity-dominated to the disassembled state also gives the particles an extra spin, thus deviating from the underlying assumption that $\acute{v}=v$ in the analytical model. 

If the planet is also rotating prior to entering the tidal sphere, these points are amplified. How can this explain the results in Figure \ref{fig:a3auI}? The cut-off in Panel \ref{fig:a0_1q3a1mI} suggests that a combination of a large planet and a deep disruption expedite the moment of breakup. Simply put, the planet disrupts prior to the moment of closest approach, which decreases the semi-major axis dispersion. In Panel \ref{fig:a1q3a1mI}, the situation is rather different. The disruption is partial, and only the outermost particles disassemble from the rest of the planet. Additionally, here the planet deforms slowly and the radial distance of these particles during breakup likely increases, in addition to obtaining an extra spin. Both of these effects increase the semi-major axis dispersion and lead to a smaller cut-off than predicted by the analytical model. Finally, Panel \ref{fig:a1q3a1mI} depicts the state at the end of the evolution, whereas the first flyby results in a much more modest dispersion (cut-off at $\sim 0.15$ AU), and only after the second flyby of the planet do we reach the final $\sim 0.05$ AU cut-off. The previously mentioned effects are most likely amplified by the newly gained fast-rotation of the planet prior to entering the tidal sphere for the second time. 

Interestingly, for Figure \ref{fig:a3auI}, the overall outcome in the semi-major axis dispersion turns out to be similar, despite three, very different disruption modes. This relatively constant outcome in varying $q$ corroborates previous results for stellar disruptions that also found an approximate constancy in the (equivalent) energy spread \citep{StoneEtAl-2013,GuillochonRamirez-Ruiz-2013,SteinbergEtAl-2019}. To our knowledge, this paper is the first to suggest the same in planetary disruptions.

Lastly, we note that none of the inner cut-offs is as small as that of the disintegrating minor planet around WD 1145+017. At $\sim 0.005$ AU \citep{VanderburgEtAl-2015}, its orbit is one-fifth as small as the smallest value in Panel \ref{fig:a0_1q3a1mI}, which could be another strong piece of evidence that this object has undergone further evolution that tightened and circularized its orbit.

Figure \ref{fig:a3auI} also features a clear asymmetry between the bound and unbound (negative semi-major axis) material, which could be easily explained by repeated disruptions. According to the analytical model, we might expect complete symmetry, however, the histogram is normalized by the number of disrupted fragments. Unbound fragments (or 'escaped' in Panel \ref{fig:a0_1q3a1mI}) never return to disrupt further, whereas bound fragments do return, and are repeatedly dissected to smaller and smaller pieces (according to the analysis in Section \ref{SS:SizeDistribution}) and thus increase in numbers. In Panel \ref{fig:a0_1q3a1mI}, the disruption is very deep and the planet breaks into its constituent particles already during the first flyby. The relatively small number of fragments that return for subsequent disruptions cause the distribution to depart from symmetry, but the effect is small. In contrast, the bound debris discs in Panels \ref{fig:a0_5q3a1mI} and \ref{fig:a1q3a1mI} are heavily processed by repeated disruptions, thus increasing fragment numbers. It is important to keep in mind that here the asymmetry is not indicative of the mass fraction of bound and unbound materials, which are approximately equal.

\begin{figure}
    \begin{center}	
	    \includegraphics[scale=0.5]{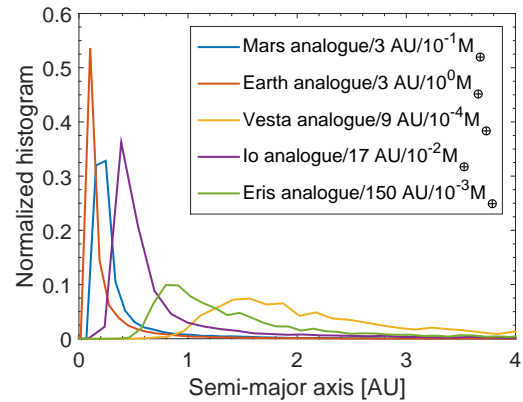}
		\caption{Semi-major axis normalized histogram of the final bound debris discs, formed by tidal disruption of several Solar system analogue planets/dwarf-planets, around a 0.6$M_\odot$ WD. All have a similarly large pericentre of $q=1R_{\odot}$. Resolution is 200K particles.}
		\label{fig:a1qall}
	\end{center}	
\end{figure}

In Figure \ref{fig:a1qall} we further show all the debris discs from Table \ref{tab:HybridModels} that have a pericentre distance of $q=1R_{\odot}$, while the semi-major axes and masses vary. In spite of varying in semi-major axis, all of these simulations approximately satisfy the condition $r_\mathrm{crit}\ll R$, hence the cut-off should be proportional to $d^2/2R$, but smaller by some factor due to the large $1R_{\odot}$ pericentre distance (as we have discussed in detail for Panel \ref{fig:a1q3a1mI}). It is easily recognized that the innermost semi-major axis cutoff and the peak in each curve correlates (left to right) with the planet size/mass. The bigger the planet, the smaller $d^2/2R$. As previously mentioned, the innermost semi-major axis cut-off points correspond to the surface constituent particles (for which $r=R$ and $d^2/2R$), however the peaks in the semi-major axis dispersion correspond to constituent particles within the planets (i.e., $r<R$). Mathematically, the distributions from peak to cut-off should therefore be narrower for the larger planets, as is clearly the case.

\subsection{Size distribution}\label{SS:SizeDistribution}
In principle the tidal disruption problem is determined from the density ratio of the planet and the star. The higher the (relative) planet density, the lower the Roche limit. If the object passes within the Roche limit the material disrupts. Setting aside other complicating aspects such as self-rotaion, intrinsic material strength and complex internal structures, if an object continues to pass within the Roche limit it should in principle continue to disrupt to smaller and smaller pieces until the latter finally reach a size in which the material strength dominates, preventing any further disruptions.

In our numerical simulations, however, the minimum size that can be attained is determined by the resolution. Even our largest resolution, which exceeds that of any previous tidal disruption study by approximately two orders of magnitude, still involves large single SPH particles of a characteristic size of at least a few km. The limiting physical size of strength-dominated particles which can withstand being ripped apart by the tidal forces is however of the orders of tens of meters to hundreds of meters \citep{BrownEtAl-2017}, and perhaps even smaller. Therefore, our initial expectation is that eventually, our hybrid simulations would results in a flat size distribution - exactly the size of a single SPH particle.

\begin{figure*}
    \begin{centering}
        \subfigure[t=0, before disruption.] {\label{fig:SizeDist0Orb}\includegraphics[scale=0.53]{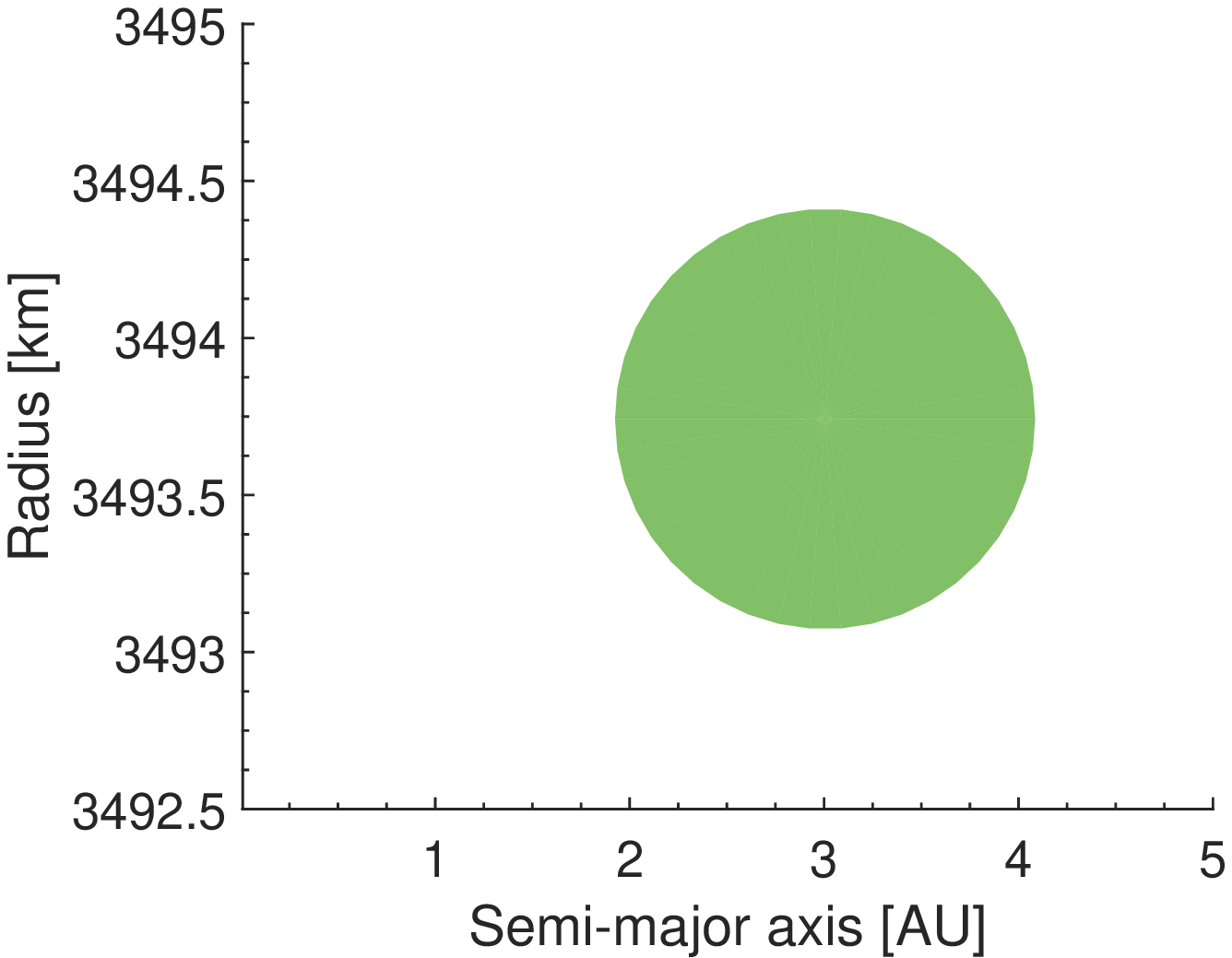}}
    	\subfigure[t=0.02 planet orbits.] {\label{fig:SizeDist0_02Orb}\includegraphics[scale=0.53]{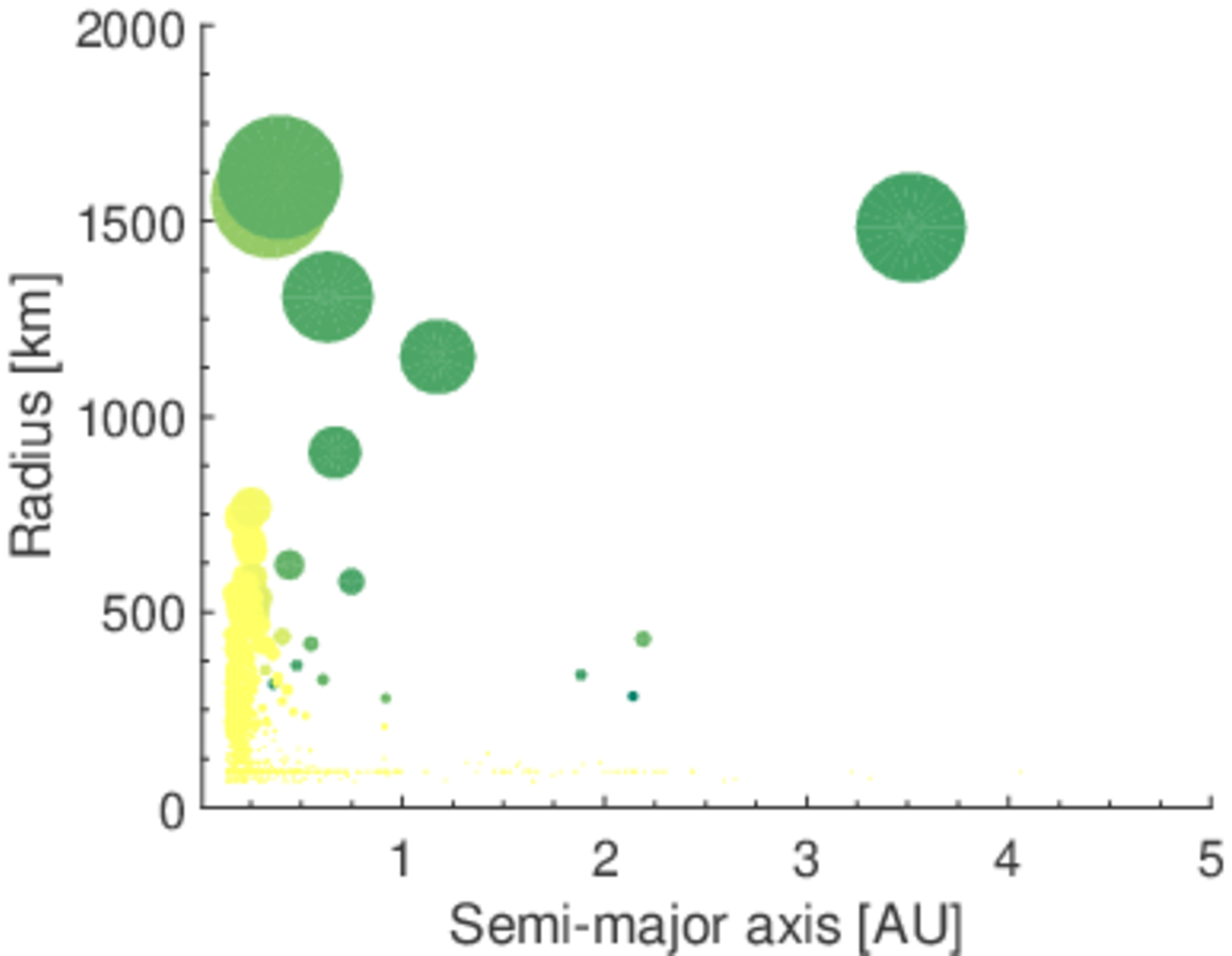}}
 		\subfigure[t=0.2 planet orbits.]
 		{\label{fig:SizeDist0_2Orb}\includegraphics[scale=0.53]{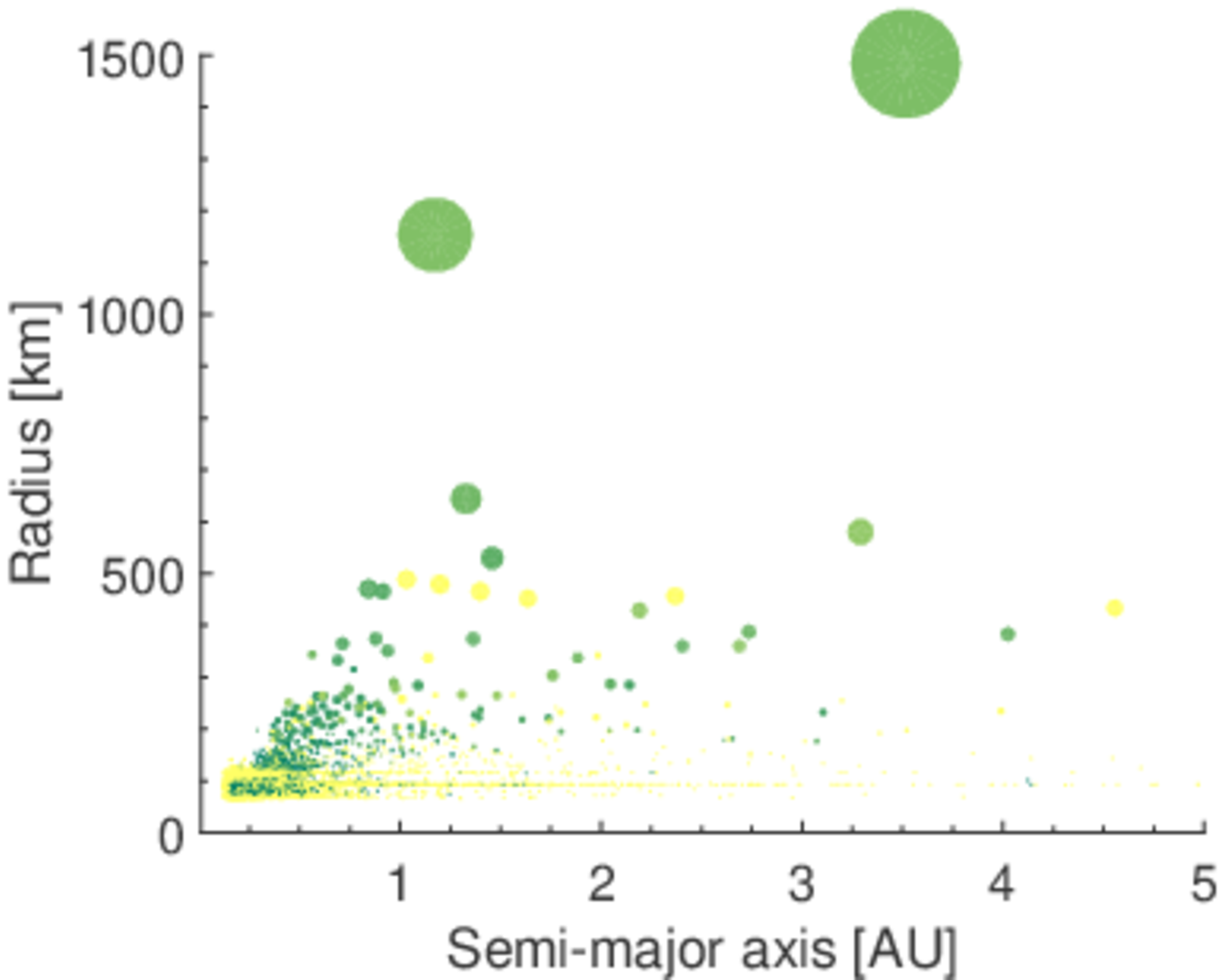}}
		\subfigure[t=1 planet orbits.] {\label{fig:SizeDist1Orb}\includegraphics[scale=0.53]{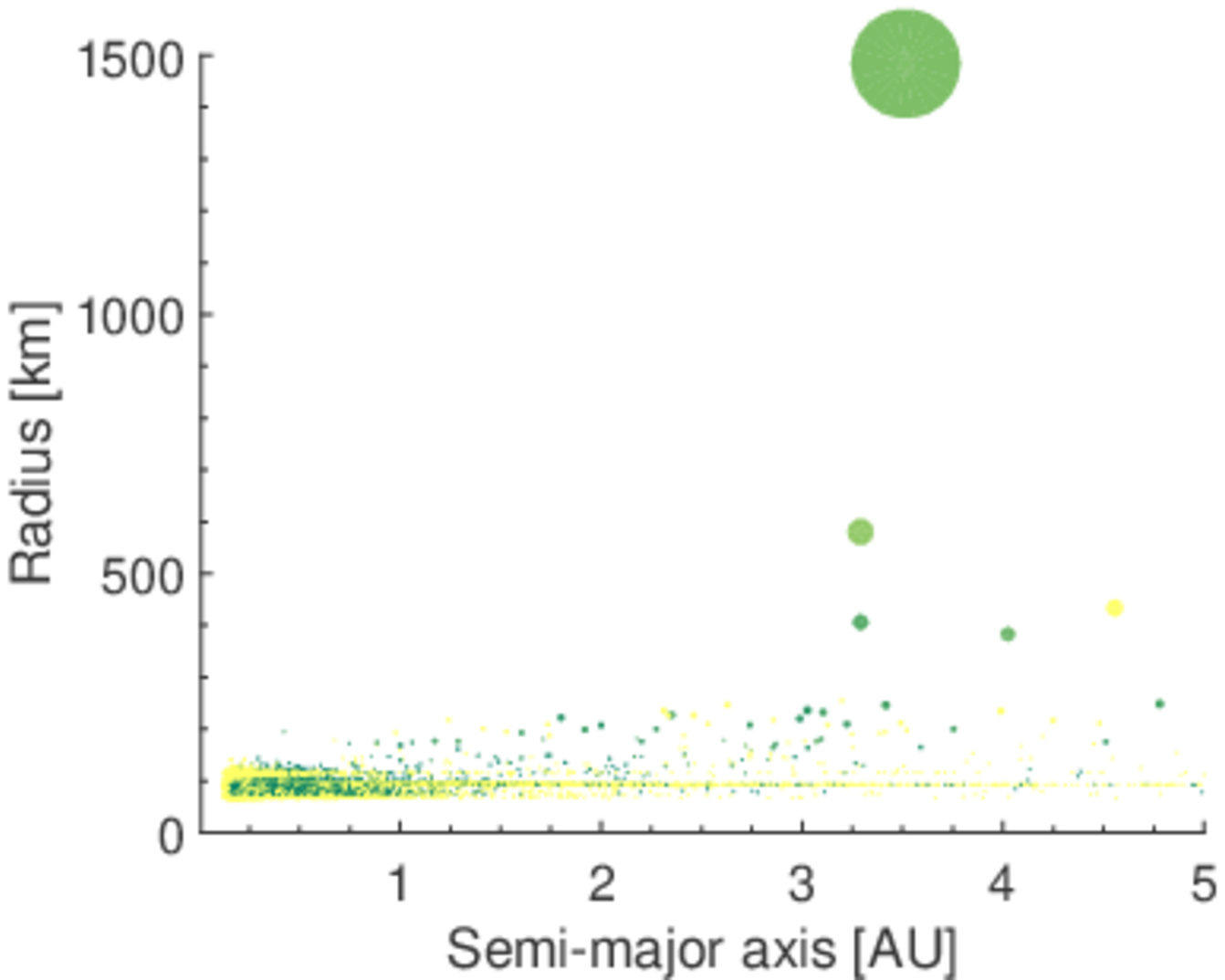}}
		\subfigure[t=2 planet orbits.] {\label{fig:SizeDist2Orb}\includegraphics[scale=0.53]{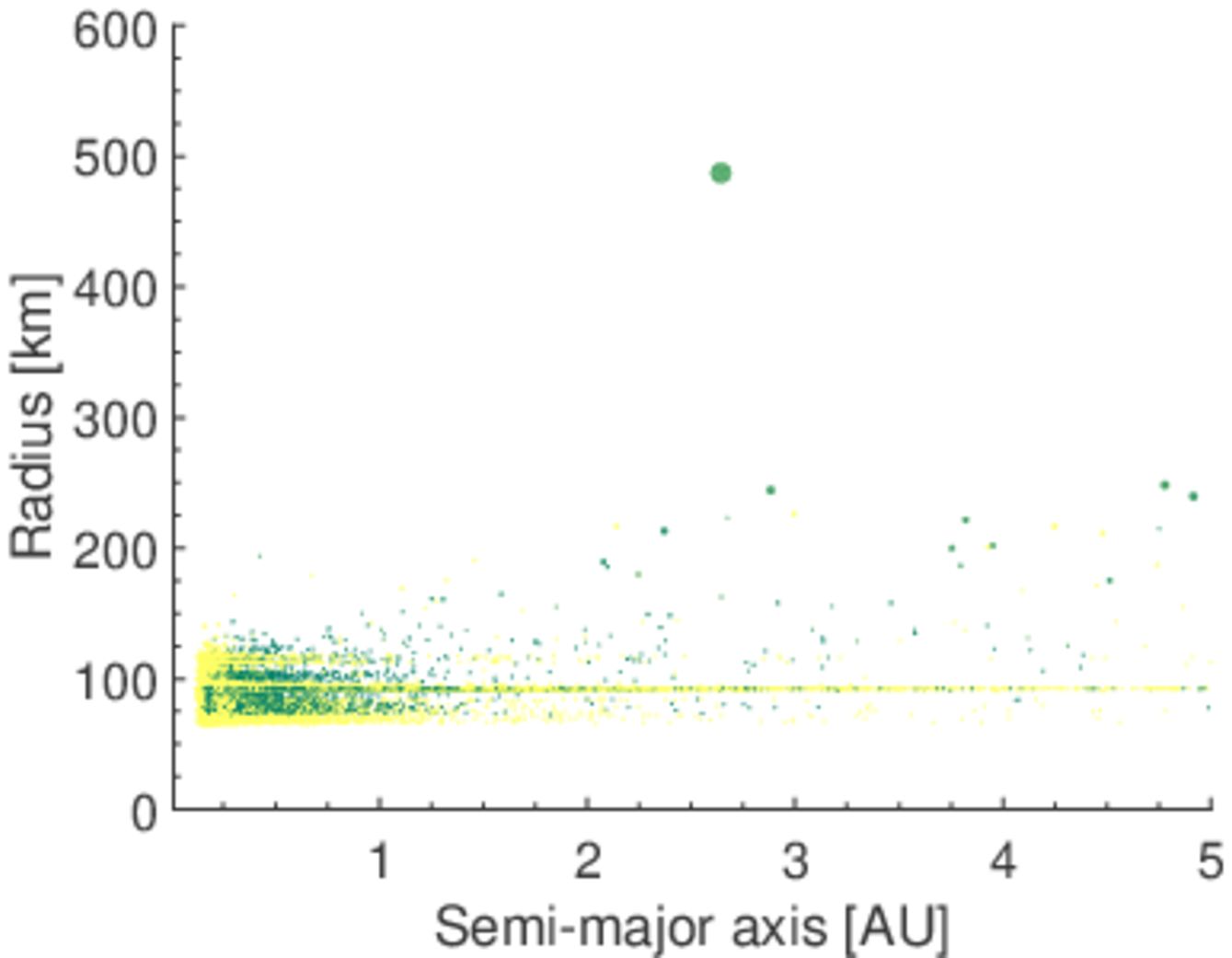}}
 		\subfigure[t=11471 planet orbits.]
 		{\label{fig:SizeDist11741Orb}\includegraphics[scale=0.53]{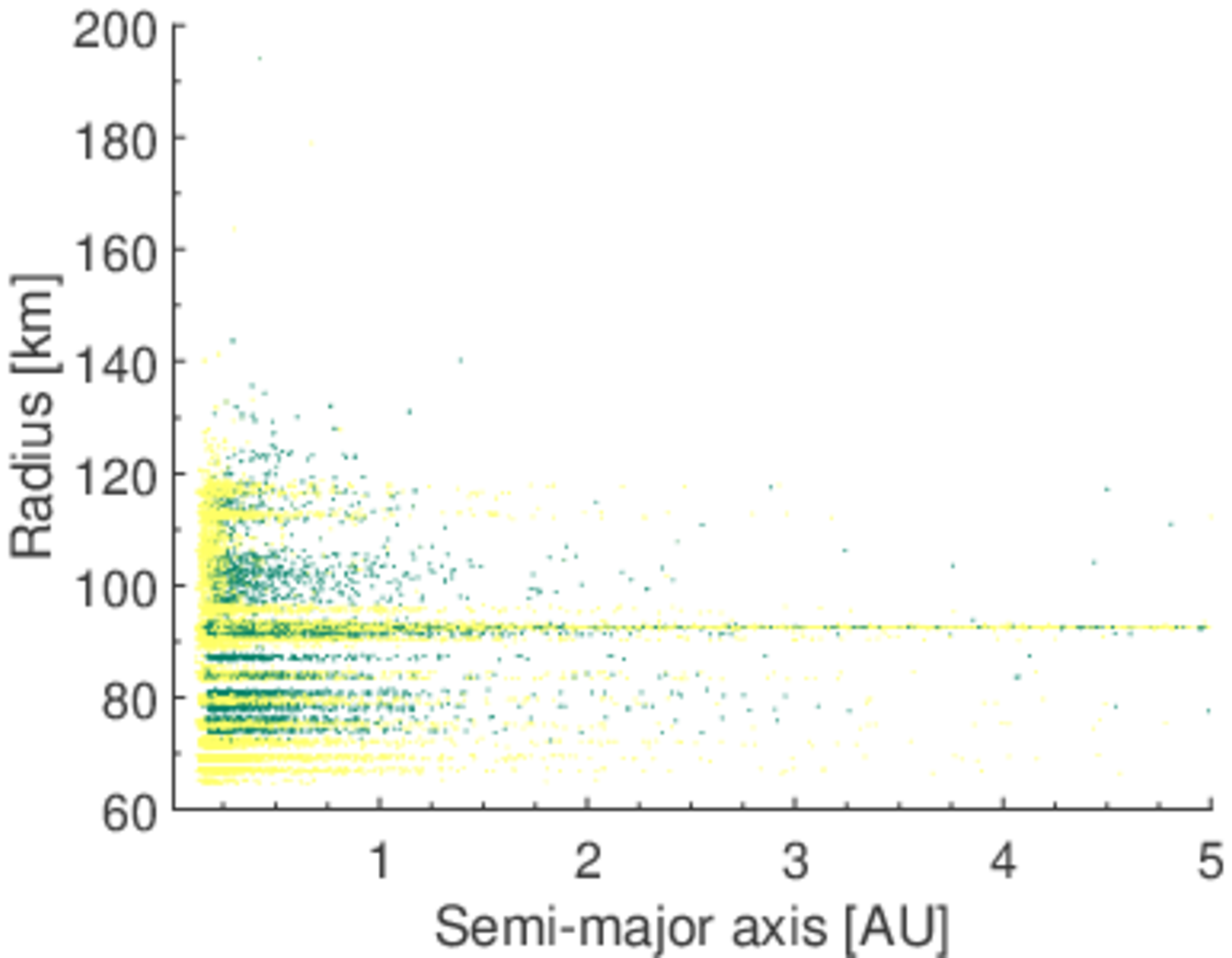}}
		\subfigure{\includegraphics[scale=0.53]{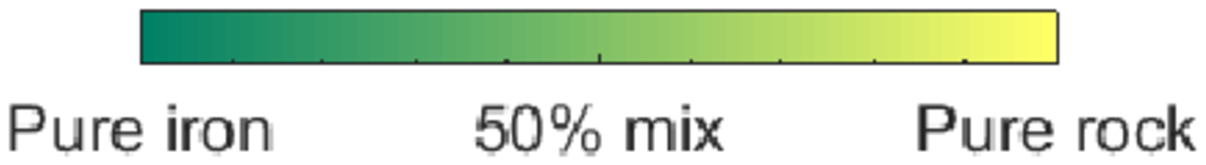}}
	\end{centering}
	\caption{Temporal evolution of fragment size distribution, of a tidally disrupted Mars analogue ($a=3$ AU and $q=0.5R_{\odot}$) around a 0.6$M_{\odot}$ WD. Fragment radii are plotted as a function of their semi-major axis. Each fragment is depicted by a circle whose size is scaled with the number of its constituent SPH particles, and the colour coding represents different compositions (see underlying colorbar). Panels (a)-(f) depict the time progress in units of the original planet's orbital period, at $a=3$ AU. After 11471 such orbital periods, the last remaining bound fragment disrupts.}
 	\label{fig:SizeDist}
\end{figure*}

In Figure \ref{fig:SizeDist} we show how the fragment size distribution of a Mars analogue at 3 AU indeed evolves towards the numerical minimum, as a function of time. Fragment radii are plotted as a function of their semi-major axis. Each fragment is depicted by a circle whose size is scaled with the number of its constituent SPH particles, and the colour coding represents different compositions, as in the underlying colorbar. In Panel \ref{fig:SizeDist0Orb} we show the original Mars-like planet, with approximately the correct dimensions, and an orbit at precisely 3 AU. Then, in Panel \ref{fig:SizeDist0_02Orb} the planet undergoes its initial tidal breakup, disrupting and then fragmenting into multiple clumps with tight orbits, in agreement to Figure \ref{fig:a1qall}, and one major bound fragment with a semi-major axis greater than that of the original planet. As we advance to the next panels there is a sequential flattening in the size distribution, proceeding from left to right. The last remaining major fragment is beyond 3 AU, and thus it takes longer than 1 orbital period of the original planet before it can disrupt. After an additional orbit time, in Panel \ref{fig:SizeDist1Orb}, the evolution is nearly complete. Full completion is however reached only when the last bound, extremely eccentric fragment, returns to the tidal sphere one last time.

Note that at the end of the simulation in Panel \ref{fig:SizeDist11741Orb}, not all the particles follow a single horizontal line, as might be expected. Instead there is one prominent line in addition to multiple others, that span a range of sizes, despite reaching the minimum single SPH particle limit. The reason for this behaviour is that the SPH particles span across a range of densities. When we inspect the data in detail, we observe that during this particular disc formation scenario many of the particles inhabit the cold expanded region in the Tillotson equation of state due to the pulling apart motion of the tidal stream, for the initial (major) tidal disruption. The cold expanded state in the Tillotson equation of state phase space is the region in which the density is lower than the reference density (at zero pressure) and the internal energy is lower than the energy of incipient evaporation. This region of the phase space is reached through tension rather than compression (see \cite{BurgerEtAl-2018}). 

We also analyse other simulations and find that low particle densities may be reached through a different route - the vaporization or partial vaporization regimes - which occur when the internal energy surpasses the energy of complete or incipient vaporization, respectively. Such high internal energy states are not reached in the scenario presented in Figure \ref{fig:SizeDist}, however for an Earth-mass planet with the same orbital parameters, we do see some particles with internal energies above the energy of incipient vaporization. These particles gain their energy during the fragmentation phase, when the stream collapses under its own gravity and the compression in the outer layers of the largest fragments is sufficiently high. High internal energies are especially dominant in the more violent tidal disruptions with $q=0.1R_{\odot}$, however in this case they occur at a much earlier stage when the planet passes through the tidal sphere.

\subsection{Formation time}\label{SS:FormationTime}
The typical outcome of tidal disruptions is to grind down a disrupted planet (bound disc) into smaller and smaller pieces. In our simulation suite, planets are large enough and orbit far from the star, and therefore fragments are often flagged to very tight orbits. It is also possible to find bound fragments that are flagged out to larger orbits than that of the original planet, however their fraction is typically very small, and depends on the exact parameters of the problem. By that logic, all the fragments interior to the original semi-major axis are expected to re-disrupt before one orbit of the original planet passes. They are obviously much smaller than the original planet, and at least initially their size follows a power law distribution (Section \ref{SS:SizeDistribution}). Given their smaller size and typically much tighter orbits their subsequent disruptions are less dispersive than that of the original planet, but they nevertheless keep dissecting into smaller pieces, since their pericentre distance remains the same. Figure \ref{fig:SizeDist} is a classic depiction of the aforementioned behaviour.

Now consider a simple exercise. A fragment disrupts into sub-fragments that follow a power-law size distribution. Further consider a limiting case wherein the largest sub-fragment is half the size of the original planet and has the same semi-major axis. In this scenario the largest fragment would reach one thousandth of the size of the original fragment, within merely 10 orbital periods ($2^{10}$). In fact, only the fragments that are initially flagged out to larger orbital periods than that of the original planet take a long time to disrupt and complete the formation of the disc. This is exactly what we see in various hybrid models (e.g. Figure \ref{fig:SizeDist11741Orb}).

By rule of thumb, therefore, the formation of the disc should be rapid and typically take only a few orbits of the original planet. This is indeed the outcome in all of our hybrid simulations. In Table \ref{tab:HybridTimescale}, the disc formation progress for our hybrid models is shown as a function of time, or rather as a function of orbital periods of the original planet. Each row corresponds to a different hybrid simulation. The name is listed in the first column, the pericentre distance in the second, and then the progress in \% denotes the ratio between the number of bound fragments at time $t$ to that at the end of the simulation.

\begin{table}
	\caption{Hybrid disc formation progress (\%).}
	\begin{tabular}{*{5}{l|}}	
		\hline
		{Hybrid model} & $q$ ($R_{\odot}$) & {1 Orbit} & {2 Orbits} & {5 Orbits} \\
		\hline
        \emph{Earth}	& 0.1 & 99.85	& 99.89 & 99.94\\
						& 0.5 & 96.91 	& 98.79 & 99.21 \\
						& 1.0 & 53.41 	& 73.99 & 91.71 \\
    	\emph{Mars}  & 0.1 & 98.87 & 99.53 & 99.89\\
 					 & 0.5 & 88.12 & 96.89 & 99.78\\
 					 & 1.0 & 10.46 & 55.43 & 98.54\\
		\emph{Main-belt (Vesta)} 	& 1.0 & 29.53 & 84.10 & 96.17\\
		\emph{Satellite (Io)} 		& 1.0 & 12.13 & 56.96 & 99.98\\
		\emph{Kuiper-belt (Eris)} 	& 1.0 & 36.70  & 74.31 & 100.0\\
		\hline
	\end{tabular}
	\label{tab:HybridTimescale}
	The first two columns show the name and simulated pericentre distance of each analogue planet. The last three columns show the disc formation progress in \%, i.e. the ratio between the ongoing number of bound fragments to the final number at the end of each simulation, as a function of the number of orbital periods (of the original planet) since the initial planet disruption.
\end{table}

In general, the rate of disc formation inversely correlates with the pericentre distance. Even in simulations in which the pericentre is the largest and close to the Roche limit, 5 orbits are sufficient in order to gain a considerable progress. In all simulations the disc formation is at least 91\% complete by merely 5 orbital periods of the original planet. We note that in Table \ref{tab:HybridTimescale} the orbital period increases with each simulated planet (see Table \ref{tab:HybridModels}), and therefore in absolute terms the progress in $t$ is slower.

\subsection{Self-rotation distribution}\label{SS:RotationExcitation}
We investigate the effect of tidal spin-up of fragments, with the caveat that as disc formation progresses, the size distribution flattens out, eventually dissecting all fragments to single SPH particles (see Section \ref{SS:SizeDistribution}). For the latter, rotation cannot be calculated, and we can only postulate on the self-rotation properties of single SPH particles by extrapolation, relying on earlier epochs in which we still have large fragments in our disc. We nevertheless attempt to quantify the fragment rotational properties \emph{during} the disc formation, because their rotation enhances tidal breakup (see related discussion in Paper I), could contribute to understanding phenomena like the Yarkovsky effect (see appendix of Paper I) and since fast-spinning fragments can be further spun up in a manner similar to that suggested by \cite{MakarovVeras-2019}, if the Roche limit is effectively reduced owing to fragment internal strength. 

\begin{figure}
	\begin{center}	
		\includegraphics[scale=0.56]{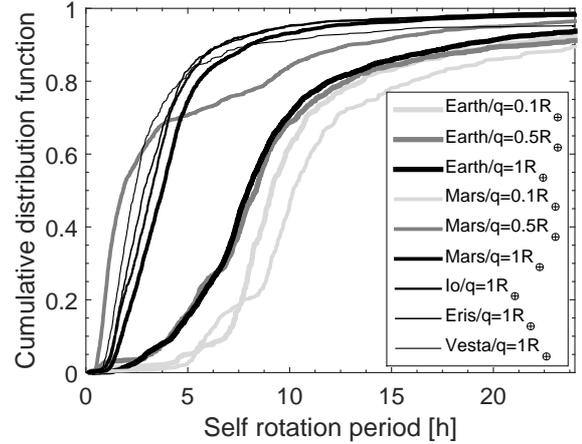}
		\caption{The cumulative distribution function of disc fragment rotation periods for all hybrid models in Table \ref{tab:HybridModels}, after 1 orbital period of the original planet. The line thickness correlates with planet mass and its colour with pericentre distance, as shown in the legend. The majority of simulations show very rapid fragment rotation.}
		\label{fig:RotationPeriods}
	\end{center}	
\end{figure}

In Figure \ref{fig:RotationPeriods} we plot the cumulative distribution function of all our hybrid simulations to gain some statistical insights. We only calculate the orbital period for fragments with two or more SPH particles, using the method in \cite{MalamudEtAl-2018}, while single SPH particles do not provide any rotational information and are therefore ignored. Note that at the beginning of disc formation, we have small number statistics for the $q=1R_{\odot}$ simulations (the disruptions being partial), while for the $q=0.1R_{\odot}$ simulations, virtually all fragments are close to the single SPH particle size, even after the first planet disruption, hence we have a similar problem. Based on Table \ref{tab:HybridTimescale}, we find 1 orbital period a suitable choice to minimize small number statistics, at least for the majority of simulations.

As indicated by the legend, the sub-plots differ such that the more massive planets are depicted by thicker lines. Additionally, the grayscale tone correlates with the pericentre distance, such that the black colour corresponds to $q=1R_{\odot}$ simulations. The results indicate two major groupings. The Earth simulations and the $q=0.1R_{\odot}$ Mars simulations are in the first group, while the rest of the simulations are in the second group. Both groups are characterized by fast rotating fragments, however in the former group some 70-80\% of the fragments have rotation periods less than merely twice the 2.2 h cohesionless asteroid spin-barrier \citep{PravecEtAl-2002}.

\subsection{Resolution dependence on convergence}\label{SS:Resolution}
We test the effect of resolution on the simulation results. Our main goal is to investigate convergence in terms of the bound disc fragment semi-major axis distribution.

\begin{figure}
    \begin{center}
	    \includegraphics[scale=0.56]{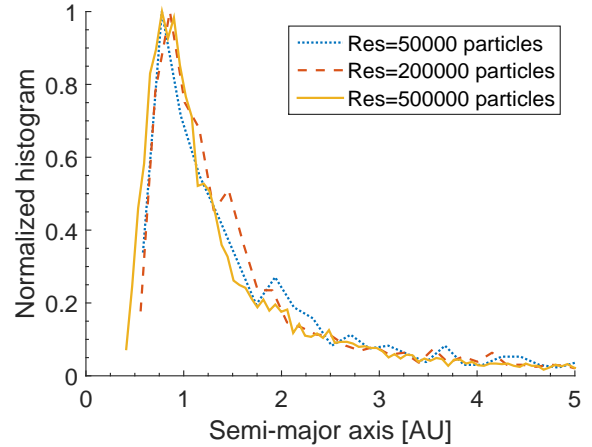}
		\caption{Resolution dependence, comparing the eventual debris disc of a tidally disrupted Kuiper-belt analogue ($a=150$ AU and $q=1R_{\odot}$) around a 0.6$M_{\odot}$ WD. Three superimposed normalized histograms for the fragment semi-major axis are presented, given low (50K/dotted line), intermediate (200K/dashed line) and high (500K/solid line) number of SPH particles.}
		\label{fig:Resolution}
    \end{center}	
\end{figure}

For our comparison we simulate the Kuiper-belt/Eris analogue case, where $a=150$ AU and $q=1R_{\odot}$, in three different resolutions. In addition to our fiducial resolution value of 200K SPH particles, we also produce two simulations with low (50K) and high (500K) resolutions. Figure \ref{fig:Resolution} then shows the normalized histograms of these three simulations (see legend), such that each histogram peaks at 1, and their morphologies can be directly compared. We see that increasing the resolution has a relatively small impact. Thus, we conclude that the main characteristic of the disc, its semi-major axis distribution of fragments, is hardly affected by resolution.

We do not check resolution convergence for the disc fragment size distribution since the latter tends to be flat anyway, due to the continuous grinding down of all the bound debris to the level of smallest resolved single particles. Nevertheless, in Appendix \ref{A:ResolutionDiscSizeDist} we show that fragments with a large pericentre distance (around $1R_{\odot}$) and a small number of SPH particles (<500) do not undergo the expected disruption and flattening of the size distribution, however this behaviour is not physical but rather results from numerical issues. We show that it can be solved either by increasing the resolution or by decreasing the values of the artificial viscosity parameters. We thus conclude that the grinding down and flattening of the size distribution remains a valid physical interpretation.

\section{Unbound material: ejection of interstellar asteroids}\label{S:Unbound}
\subsection{Size distribution}\label{SS:UnboundSizeDistribution}
As already mentioned, the scenarios in Table \ref{tab:HybridModels} are all characterized by typical $r_\mathrm{crit}$ values much smaller than the object radius, which means that nearly half of the initial planet mass becomes unbound. The latter particles/fragments are ejected to the interstellar space, hence, unlike bound debris, they are not repeatedly dissected by subsequent disruptions to flatten out their size distribution. Instead, fragment sizes depend on the pericentre/breakup distance (see also Section \ref{SS:UnboundRotationDistribution}). In all but the deepest disruptions, the emerging streams collapse under their own self-gravity to form large fragments from the debris. We thus expect a non-flat, and perhaps a power-law size distribution. In contrast, very deep disruptions violently break the planets into small, typically single SPH particles, and the emerging streams of debris do not collapse gravitationally to form larger fragments (see Section \ref{S:Intro}), hence we expect an outcome more closely resembling a flat distribution, with only a small number of fragments that coalesce from the denser part of the stream.

\begin{figure}
	\begin{center}
		\includegraphics[scale=0.56]{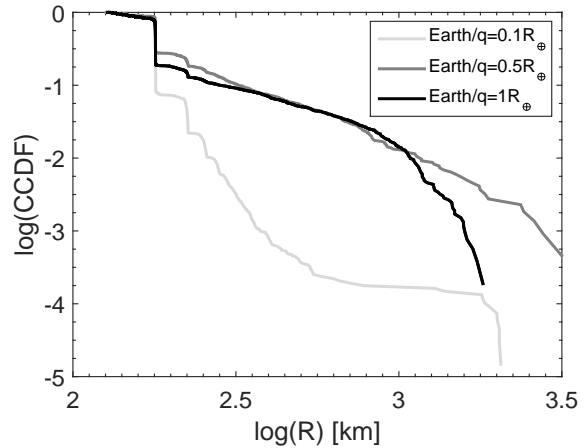}
    	\caption{Comparing the size distribution of unbound material from a tidally disrupted Earth analogue ($a=3$ AU) around a 0.6$M_{\odot}$ WD, with varying pericentre distances ($q=0.1-1 R_{\odot}$, correlating with greyscale tone). The complementary cumulative distribution function is plotted as a function of size. Both axes are in logarithmic scale, and only the simulation with $q=0.5R_{\odot}$ approximately curves out a power law distribution (linear correlation in log).}
		\label{fig:UnboundSizeDist}
	\end{center}	
\end{figure}

Indeed, Figure \ref{fig:UnboundSizeDist} shows the simulation outcomes to broadly agree with these predictions. Here we compare the unbound fragments that are ejected from a tidally disrupted Earth analogue around a 0.6$M_{\odot}$ WD, for varying pericentre distances. The analysis is performed at the end of each simulation. Due to the similarity between the Earth and Mars cases (see Section \ref{SS:UnboundRotationDistribution}), only the former is plotted here.

\begin{figure*}
    \begin{centering}		
	    \subfigure[Earth analougue, $q=0.1R_{\odot}$.] {\label{fig:UnboundRotDistEarth0_1}\includegraphics[scale=0.53]{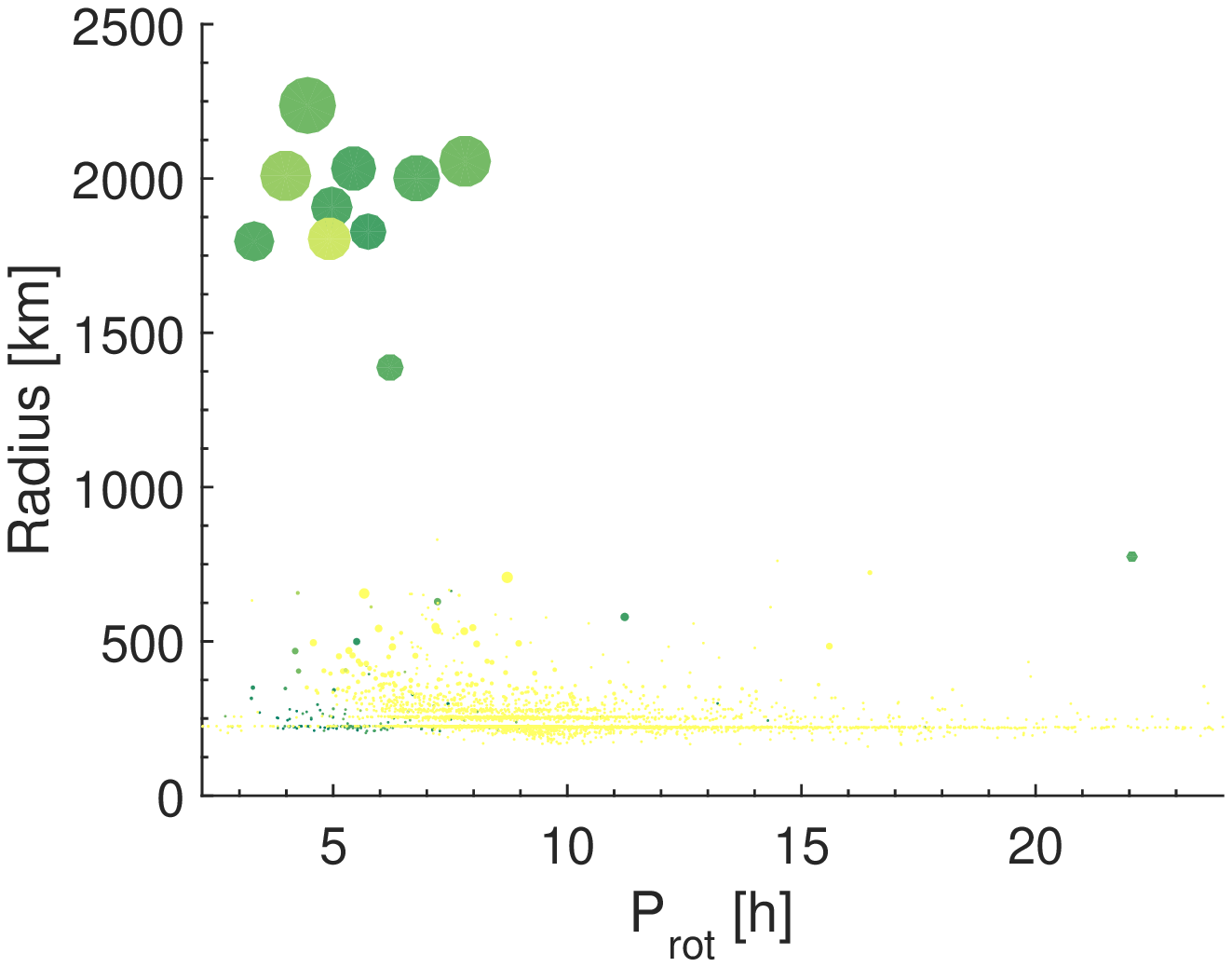}}
		\subfigure[Earth analougue, $q=0.5R_{\odot}$.] {\label{fig:UnboundRotDistEarth0_5}\includegraphics[scale=0.53]{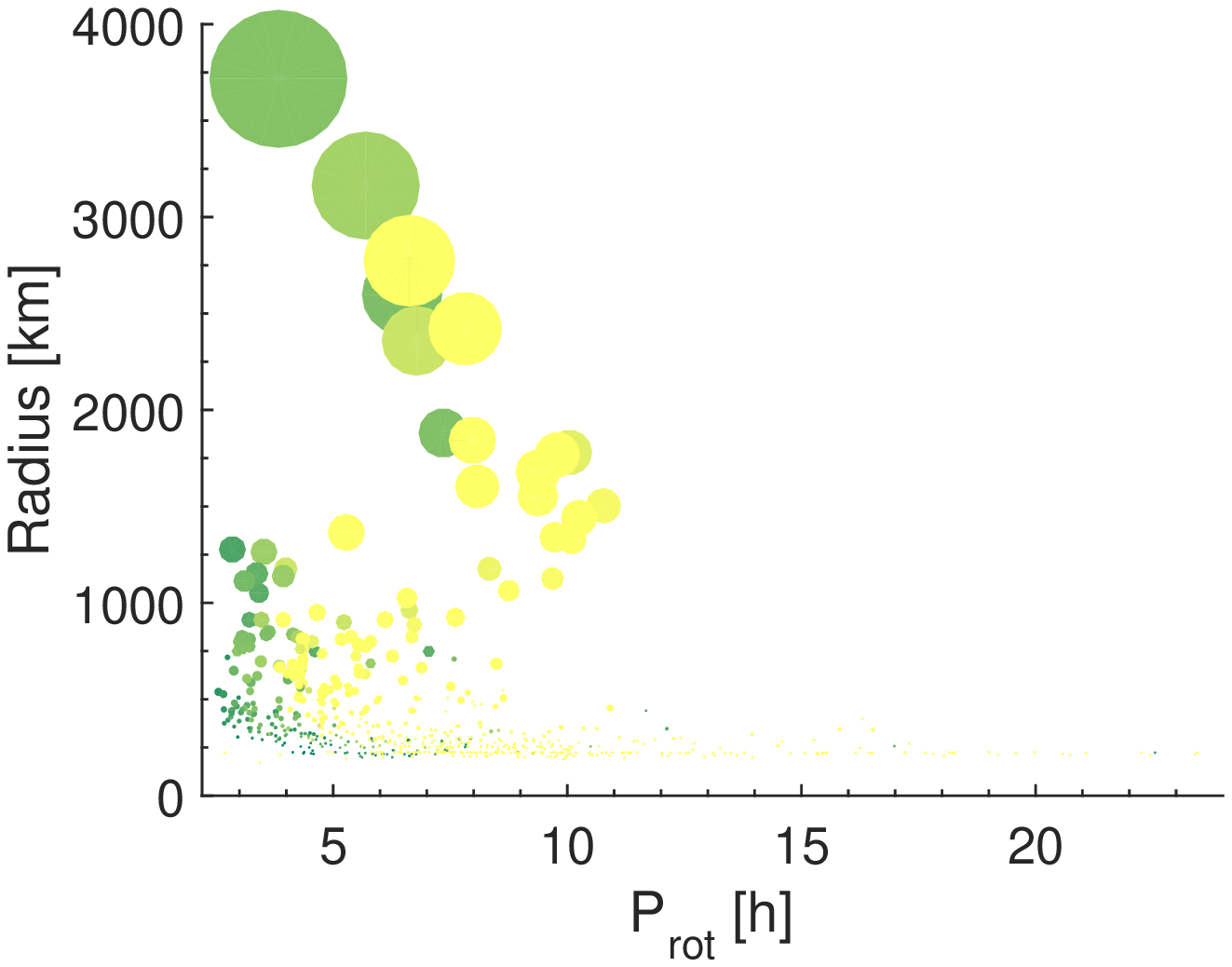}}
		\subfigure[Earth analougue, $q=1R_{\odot}$.]
		{\label{fig:UnboundRotDistEarth1}\includegraphics[scale=0.53]{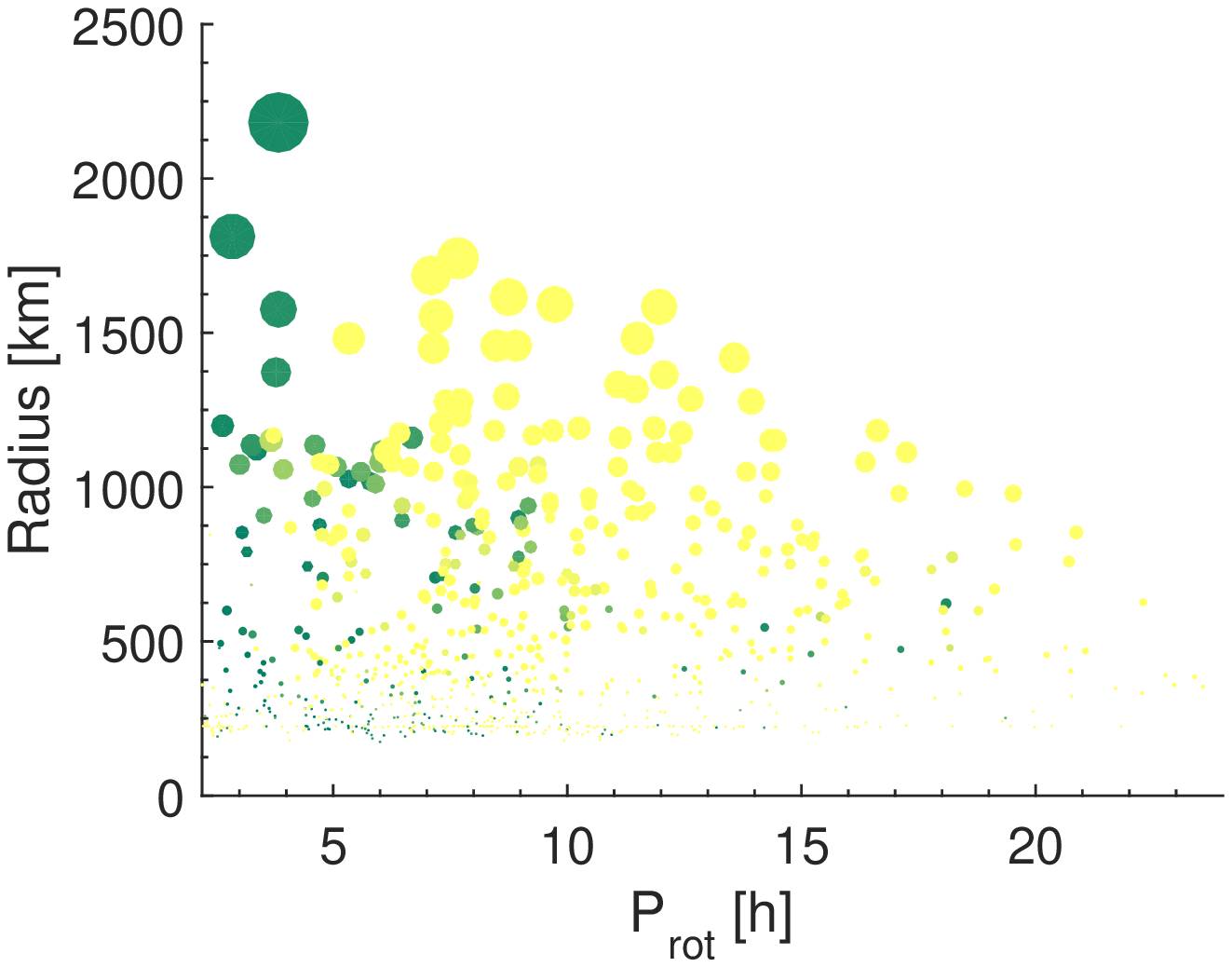}}
		\subfigure[Mars analougue, $q=0.1R_{\odot}$.] {\label{fig:UnboundRotDistMars0_1}\includegraphics[scale=0.53]{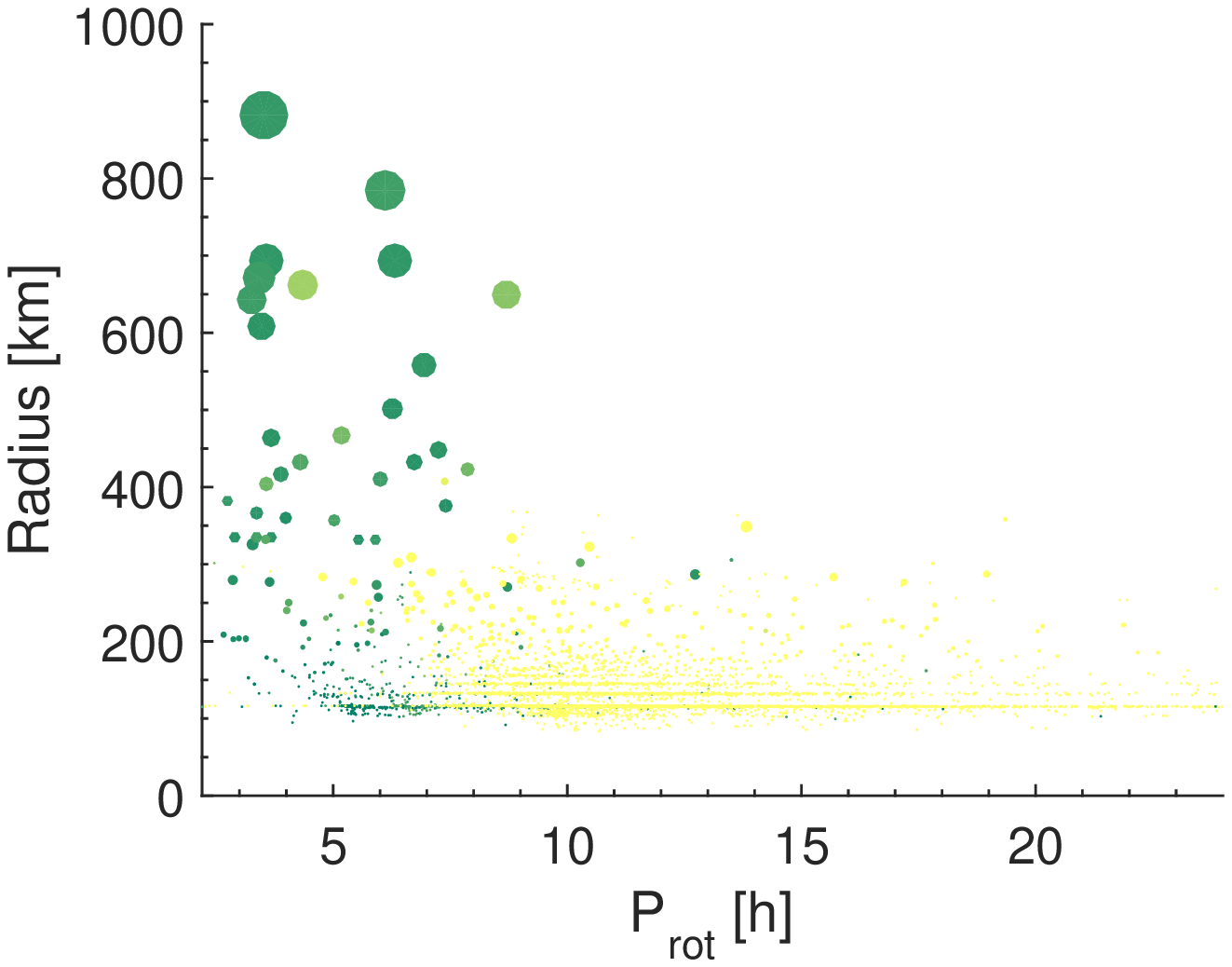}}
		\subfigure[Mars analougue, $q=0.5R_{\odot}$.] {\label{fig:UnboundRotDistMars0_5}\includegraphics[scale=0.53]{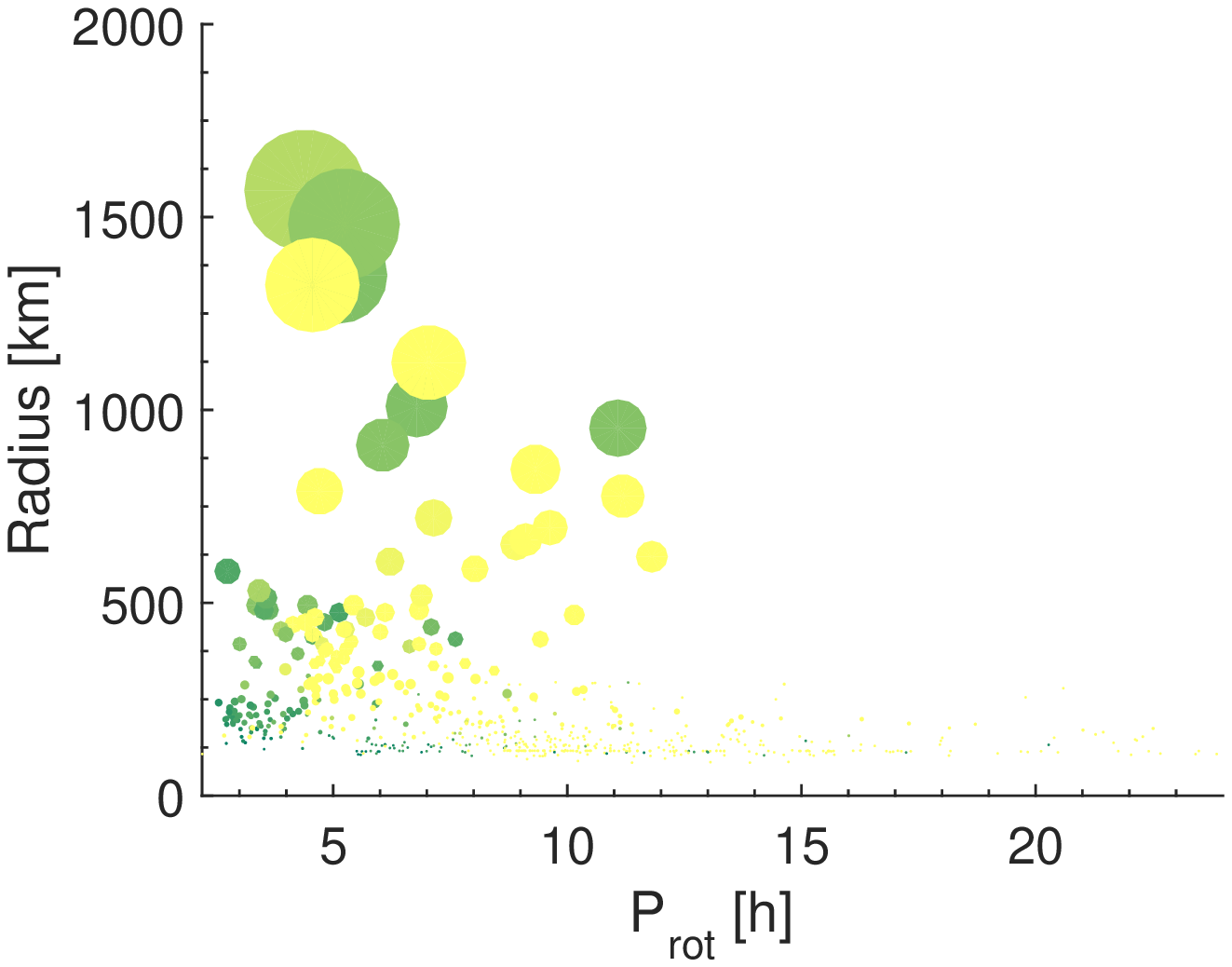}}
		\subfigure[Mars analougue, $q=1R_{\odot}$.]
		{\label{fig:UnboundRotDistMars1}\includegraphics[scale=0.53]{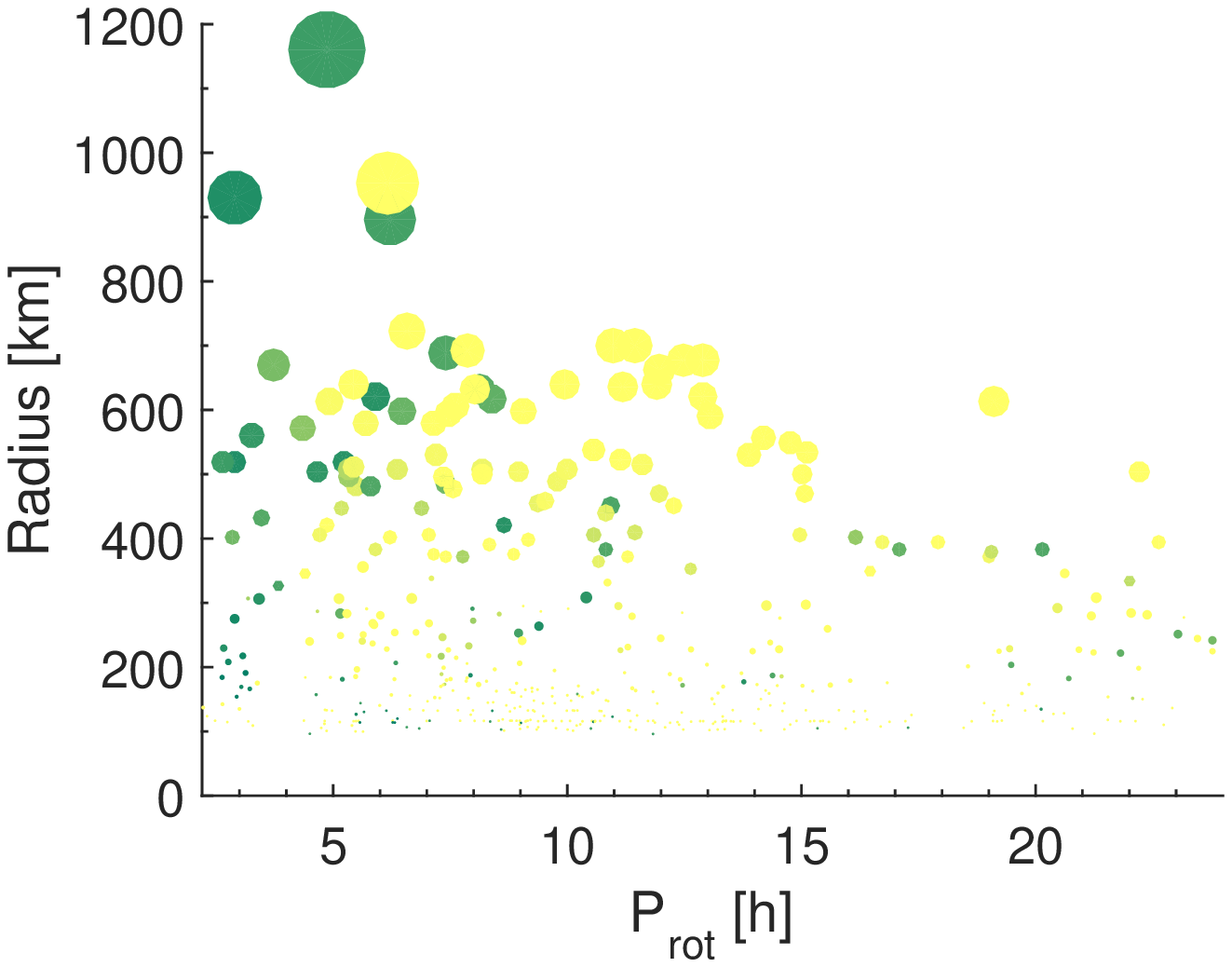}}
		\subfigure{\includegraphics[scale=0.53]{ResLegend.eps}}
    \end{centering}
	\caption{Unbound fragment self-rotation distribution, of a tidally disrupted Earth (top row) and Mars (bottom row) analogues with $a=3$ AU, and $q=0.1R_{\odot}$ (first column), $q=0.5R_{\odot}$ (second column) and $q=1R_{\odot}$ (third column), around a 0.6$M_{\odot}$ WD. The Fragment radii are plotted as a function of their rotation period. Each fragment is depicted by a circle whose size is scaled with the number of its constituent SPH particles, and the colour coding represents different compositions (see underlying colorbar). After 11471 such orbital periods, the last remaining bound fragment disrupts.}
	\label{fig:UnboundRotDist}
\end{figure*}

The complementary cumulative distribution function (or CCDF, the probability that fragment size is larger than $x$) is indicative of a power law distribution when the axes are linearly correlated (both axes are plotted in logarithmic scale). Figure \ref{fig:UnboundSizeDist} shows that only the simulation with $q=0.5R_{\odot}$ generates an approximately power law distribution for the unbound fragments. A larger pericentre distance results in an almost identical distribution for fragments up to $10^3$ km, however, larger fragments are more depleted compared to the former (See also Panels \ref{fig:UnboundRotDistEarth0_5} and \ref{fig:UnboundRotDistEarth1}). This could be explained by the fact that for $q=0.5R_{\odot}$, the first flyby results in a full disruption, ejecting most of the unbound material. In contrast, for $q=1R_{\odot}$ the first few disruptions are partial, gradually ejecting smaller fragments from the rocky mantle first, and only later on the (much smaller) iron core disrupts entirely (see also Paper I for the discussion of a similar case, with full SPH simulation). Finally, for the $q=0.1R_{\odot}$ simulation we indeed get a much flatter size distribution, as expected.

In our simulation suite, only the Earth and Mars analogue cases have simulations with $q=0.5R_{\odot}$. From these cases we can extract the unbound fragments' size distribution power law exponent (the CCDF slope subtracted by 1), which in both cases comes to approximately 4/3.

\subsection{Self-rotation and composition distribution}\label{SS:UnboundRotationDistribution}
Unbound fragments are ejected to interstellar space at high speeds, and maintain their rotational properties, as they never again interact with their star of origin. In small, Oumuamua-like fragments \citep{MeechEtAl-2017}, with irregular shapes, the angular momentum following a tidal disruption can be misaligned with the axis of maximum moment of inertia, so that the rotational velocity vector begins to precess and nutate \citep{ScheeresEtAl-2000}. Such a tumbling motion has indeed been identified for Oumuamua, and given a monolithic, high-rigidity structure, its extremely long damping timescale to return to principle-axis rotation could have kept it that way \citep{FraserEtAl-2018}. In contrast, resolution limitations in our study and the initial large size of the objects in Table \ref{tab:HybridModels}, generate much larger fragments, ranging from large asteroids to dwarf-planet size. Our goal is therefore to study the rotation properties of the so called rogue dwarf-planets/asteroids that are flung out to interstellar space.

Figure \ref{fig:UnboundRotDist} is identical to Figure \ref{fig:SizeDist}, except that its x-axis denotes the fragment's orbital period, ranging from the cohesionless asteroid spin-barrier of 2.2 h \citep{PravecEtAl-2002} to 24 h. As before, each unbound fragment is depicted by a circle whose size is scaled with the number of its constituent SPH particles, and the colour coding represents different compositions, as in the underlying colorbar. Therefore, this figure contains rotational, compositional and size information for all unbound fragments, obtained after the tidal disruption of Earth and Mars analogue planets around a 0.6$M_{\odot}$ WD, at the end of the simulation. Only these two planets have three different pericentre distances, which enables a more complete comparison.

From Figure \ref{fig:UnboundRotDist} we draw several conclusions. (a) Unbound fragment properties, including their rotation periods, composition and relative (to-progenitor) sizes are very similar between the Earth (top row) and Mars (bottom row) cases, so at least for the unbound fragments this problem appears relatively scale-invariant. (b) Between various pericentre distances (columns), however, variations are easily noticeable. Physical differences in fragment formation and in the timing of their formation (see Section \ref{SS:UnboundSizeDistribution}) completely change their properties. (c) In all panels, rotation and maximum fragment size inversely correlate. (d) Iron-composed (or mixtures with higher-than-rock bulk density) fragments generally rotate faster than purely rocky ones. The latter could be understood from breakup-velocity arguments. The breakup velocity is calculated from equating self-rotation and self-gravity, and we may express the fragment rotation period as a function of its density, such that $P_\mathrm{rot}=\sqrt{3\pi/G\rho}$, hence the inverse correlation.

\subsection{Velocity distribution}\label{SS:UnboundVelocityDistribution}
The velocity distribution of ejected planetesimals is also of particular interest in regards to the recently observed interstellar objects going through the Solar system \citep{MeechEtAl-2017,DeLeonEtAl-2019} or meteors of interstellar origin \citep{SirajLoeb-2019}. Moreover, such ejections could give rise to high velocity interstellar asteroids (or comets, although we do not consider icy compositions in this paper specifically), and their discovery can potentially pinpoint to such tidal disruption origin. We nevertheless note that unbound fragments produced in this paper, are larger than the two interstellar objects discovered in our Solar system thus far. Their small size can only be understood in the context of very deep tidal disruptions, or as second-generation fragments from intra-collisions in the unbound tidal stream (see related discussion in Paper I).

\begin{figure}
	\begin{center}
		\includegraphics[scale=0.56]{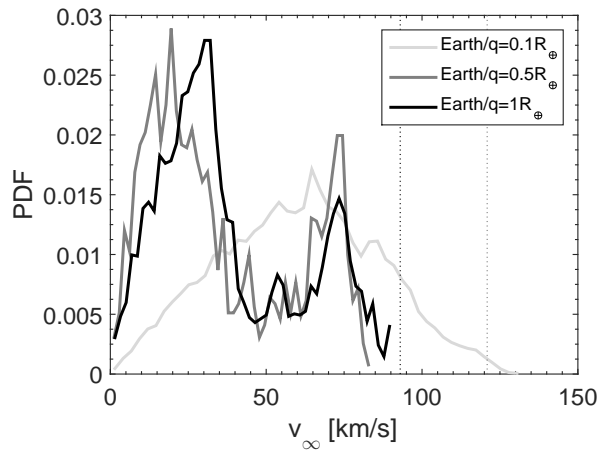}
		\caption{Comparing $V_{\infty}$ of unbound material from a tidally disrupted Earth analogue ($a=3$ AU) around a 0.6$M_{\odot}$ WD, with varying pericentre distances ($q=0.1-1 R_{\odot}$, correlating with greyscale tone). The probability density function is plotted as a function of velocity. The maximal ejection velocities from analytical predictions are shown by dotted vertical lines (extremely deep disruptions increase the velocity by $\sim$30\% and shown in grey colour).}
		\label{fig:UnboundV_Inf_Earth}
	\end{center}	
\end{figure}

\begin{figure}
	\begin{center}
		\includegraphics[scale=0.56]{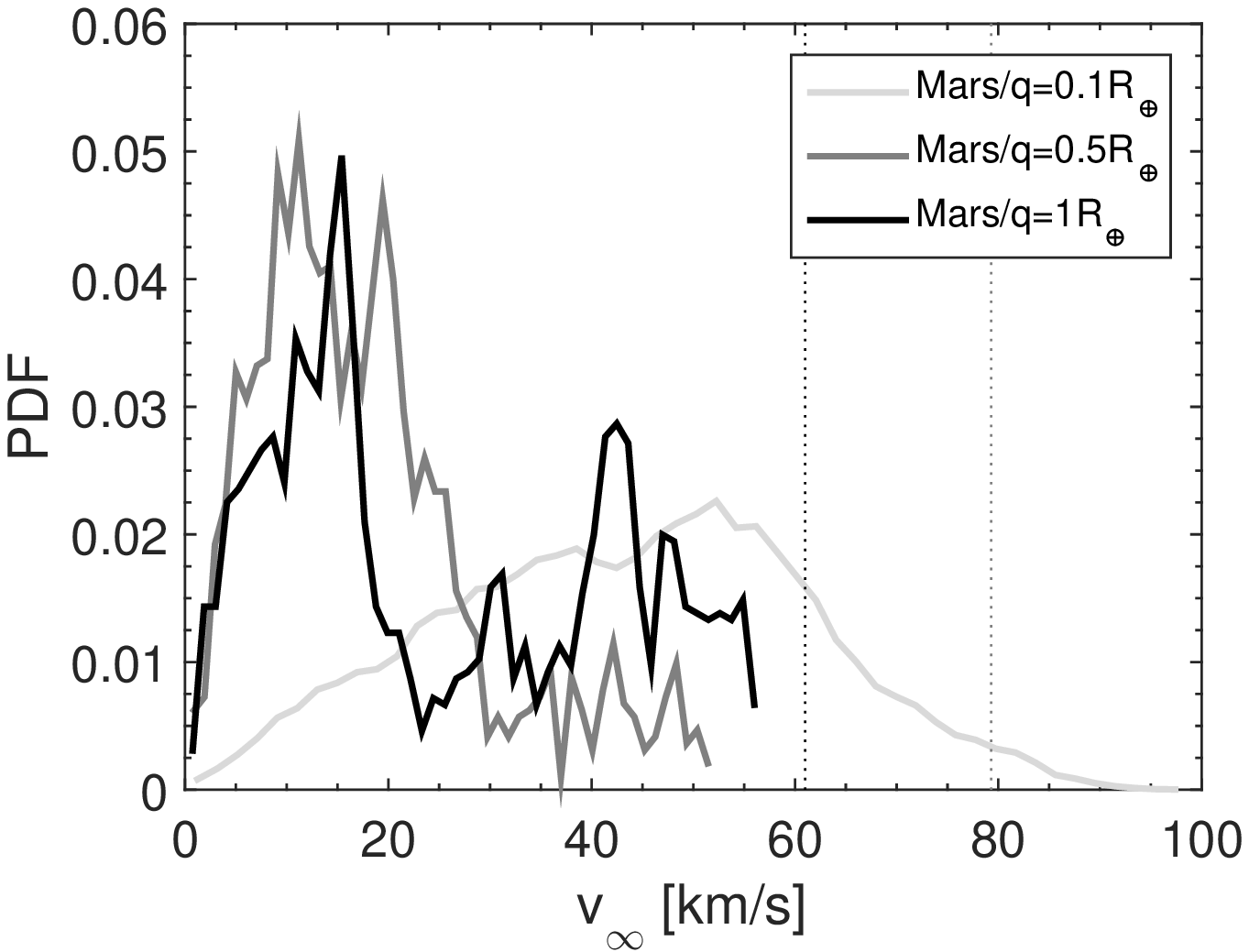}
		\caption{Comparing $V_{\infty}$ of unbound material from a tidally disrupted Mars analogue ($a=3$ AU) around a 0.6$M_{\odot}$ WD, with varying pericentre distances ($q=0.1-1 R_{\odot}$, correlating with greyscale tone). The probability density function is plotted as a function of velocity. The maximal ejection velocities from analytical predictions are shown by dotted vertical lines (extremely deep disruptions increase the velocity by $\sim$30\% and shown in grey colour).}
		\label{fig:UnboundV_Inf_Mars}
	\end{center}	
\end{figure}

\begin{figure}
	\begin{center}
		\includegraphics[scale=0.56]{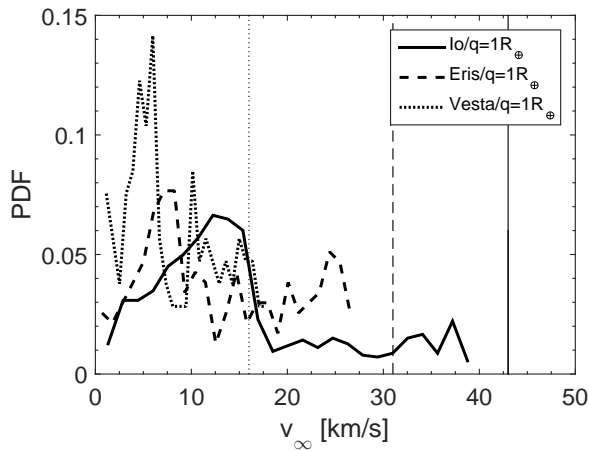}
		\caption{Comparing $V_{\infty}$ of unbound material from a tidally disrupted Io ($a=17$ AU, solid line), Eris ($a=150$ AU, dashed line) and Vesta ($a=9$ AU, dotted line) analogues around a 0.6$M_{\odot}$ WD, with pericentre distance $q=1 R_{\odot}$. The probability density function is plotted as a function of velocity. The maximal ejection velocities from analytical predictions are shown by same-style vertical lines.}
		\label{fig:UnboundSV_Inf_IoErisVesta}
	\end{center}	
\end{figure}

In Figures \ref{fig:UnboundV_Inf_Earth}, \ref{fig:UnboundV_Inf_Mars} and \ref{fig:UnboundSV_Inf_IoErisVesta} we show the detailed velocity distribution found for the ejected debris. As can be seen, ejection velocities could be highest for debris ejected from the disruption of the largest objects.

In particular, similar to the case of the ejection of hypervelocity stars following the tidal disruption of a binary star by a massive black hole \citep{YuTremaine-2003}, the ejection velocity should be of the order of:

\begin{equation}
v_\mathrm{eject}=\sqrt{2v\delta v} \label{eq:v_kick_max}
\end{equation}

\noindent where $v$ is the velocity of the disrupted planet at the tidal breakup radius, and $\delta v$ is the extra velocity kick imparted to an ejected unbound fragment. Such $\delta v$ could be at most the escape velocity for the disrupted planet. For example, envision a fragment which is just marginally bound to the planet, or think about a maximally rotating planet - the velocity of the marginally bound fragment around the center of mass of the planet is of the order of the escape velocity from that planet. Upon disruption the fragment velocity adds as a kick to the total velocity during a disruption, hence giving rise to Eq. \ref{eq:v_kick_max} following the same argument in \cite{YuTremaine-2003}. 

Given the escape velocities of Earth, Mars, Io, Eris and Vesta (11.20, 5.00, 2.53, 1.38, 0.36 km $\times$ s$^{-1}$, respectively), their densities (5.50, 3.92, 3.53, 2.52, 3.42 g $\times$ cm$^{-3}$, respectively) and taking the upper range of the likely tidal radius $r_\mathrm{t}$ ($C=1.89$, e.g. \cite{BearSoker-2013}):
\begin{equation}
\frac{r_\mathrm{t}}{R_{\odot}}=C\left(\frac{M_\mathrm{WD}}{0.6\rm M_{\odot}}\right)^{1/3}\left(\frac{\rho}{3~g\times cm^{-3}}\right)^{-1/3}\label{eq:rt}
\end{equation}
one would then expect maximal ejection velocities of 93, 61, 43, 31 and 16 km $\times$ s$^{-1}$ respectively, nicely consistent with the maximal ejection velocities we find from the detailed simulations for encounters at $q=1\,R_\odot$. Note that even higher velocities can be attained at significantly deeper encounters; as discussed by \cite{SariEtAl-2010}, these can provide for ejection velocities even up to $\sim 30\%$ higher than those obtained from Eq. \ref{eq:v_kick_max}, consistent with our $q=0.1\,R_\odot$ simulations. 

Planetesimals ejected at the lowest velocities of less than a few km $\times$ s$^{-1}$, could be recaptured again into other planetary systems (see e.g. \cite{GrishinEtAl-2019,GrishinVeras-2019}) while ejection velocities $>\sim60$ km $\times$ s$^{-1}$ are significantly higher than would be expected for debris ejected from stars in the galactic disk, and could therefore potentially point to a disruption origin, although high velocities could also be related to halo stars. An observed interstellar object with a high velocity whose direction could be traced to reside in the Galactic disk, would provide a potential smoking gun signature.

\section{Possible implications for the origin of Tabby's star, ZTF J0139+5245 and SDSS J1228+1040}\label{S:Future}

\subsection{Tabby's star - an ongoing disruption by an icy dwarf-planet?}
\label{SS:Tabys}
KIC 8462852 (a.k.a Tabby's star) is an F type star which features extraordinary dips in its light curve, of up to 20\%, first identified through the Kepler mission \citep{BoyajianEtAl-2016}. Transit events last from a fraction of a day to several days, and the observed morphology of the dips vary considerably from case to case. Later observations suggested a periodicity of 4.31 yr can be compatible with some of the observed transits \citep{SaccoEtAl-2018}. The star lacks significant excess infrared emmision \citep{ThompsonEtAl-2016}. To further the enigma of KIC8462852, \cite{Schaefer-2016} found evidence for secular dimming, with a 16\% decrease in brightness from 1890 to 1980. The Kepler data also showed similar evidence for long-term dimming, with 3\% over a 4 yr timescale \citep{MontetSimon-2016}.

Out of a multitude of hypotheses originally proposed \citep{BoyajianEtAl-2016}, a favourite among several authors \citep{BodmanQuillen-2016,NeslusanBudaj-2017,WyattEtAl-2018} was an exo-comet origin. This scenario is deemed as potentially the most likely suggestion, although it is not trivial to conceive an origin for swarms \citep{BodmanQuillen-2016} or larger clumps \citep{NeslusanBudaj-2017} of comet-like material, while also explaining the secular dimming. \cite{WyattEtAl-2018} suggest that the secular dimming might be explained by dusty material that is unevenly distributed around a single elliptical orbit, but they do not consider in detail the origin of this elliptical dust structure. 

We suggest a potential solution perfectly compatible with our hybrid model (but considering a main-sequence star instead of a WD), and can self-consistently explain the origin of the comets as well as the secular dimming. Previous studies \citep{BodmanQuillen-2016,NeslusanBudaj-2017} suggest that the mass of transiting material is around $10^{-2}-10^{-4} M_{\oplus}$. A tidally disrupted parent object with an originally eccentric orbit and compatible mass, could tidally disrupt and form smaller sub-fragments of active nuclei, spewing gas and dust. An original orbit of several AU is consistent with a non-dispersive disruption, given $r_\mathrm{crit}$ of order $\sim 10^3$ km, so it might generate a ring-like debris disc. An original orbit of several tens of AU is consistent with a bi-modal disruption, given $r_\mathrm{crit}$ of order $\sim 10^1$ km, so it might generate a dispersed disc of fragments, peaking at a few AU (as in the Vesta curve in Figure \ref{fig:a1qall}). Both configurations are therefore potentially compatible with a swarm of eccentric fragments concentrated around 3 AU (for which the corresponding orbit is tentatively around the 4.31 yr periodicity reported by \cite{SaccoEtAl-2018}).

The only caveat is that the density of the original parent object cannot be too large, otherwise it would not tidally disrupt at distances larger than the star's circumference. Additionally, note that the typical separations between Roche radius and star circumference are much lower for main sequence stars than they are for WDs. Another possibility is partial disruption of only the outer layers of a differentiated object (like in \cite{VerasEtAl-2017}). Thermo-physical models and observations from the Solar system predict such differentiated structures to be common, and the outermost layers could be composed of a mixture of ice and rock (e.g. see \cite{MalamudPrialnik-2015,MalamudEtAl-2017}). 

Finally, the secular dimming is a natural consequence of the hybrid model, as the number density of fragments increases as a function of time. While swarms of larger objects produce the large transit events, the overall flux of smaller debris in the disc increases on a timescale that is proportional to the orbit of the major fragments/original planet, and produces a gradual dimming of the star in a quasi steady-state like manner.

The detailed study of this possibility will be explored elsewhere, through the investigation of various plausible scenarios along the previously mentioned lines. We hypothesize that this explanation could be explored as a general solution for an entire class of similar dipper stars. \cite{Schaefer-2019} shows that the other dippers do not exhibit evidence for secular dimming, however, they also display much shallower transits. It could be that KIC 8462852 simply captures the phenomena at a much earlier stage.

\subsection{ZTF J0139+5245 - an early stage asteroid disruption around a white dwarf?}\label{SS:ZTF-object}
In a recent paper, \cite{VanderboschEtAl-2019} identify the second WD for which repeated transits have been observed. In contrast to the more widely known WD 1145+017, here the transits duration are much longer, of the order of several weeks, and the inferred periodicity is around 110 days.

According to the analysis of \cite{VanderboschEtAl-2019}, the likely distance of closest approach could imply an eccentricity larger than 0.97, and a maximum orbital separation of 0.72 AU. At periastron, a transiting grain would take $\sim$1 minute to transit, and as much as 1-2 hours if at apastron. The much longer transits observed, imply the presence of an extended stream of disrupted debris. 

Based on our current study, we deem the possibility of a bi-modal type tidal disruption  less likely than that of a non-dispersive tidal disruption of a small asteroid. The latter is expected to generate a coherent ring of debris with well-separated transits, as seems to be the case here, whereas the former might involve a multitude of fragments with very different timings (like in Figure \ref{fig:a1qall}). The transits duration are also compatible with a non-dispersive ring of debris, since numerical simulations in both the \cite{VerasEtAl-2014} study and in our validation tests in Paper I clearly show the gradual lengthening of the tidal stream prior to completely filling out the ring. 

If this interpretation is correct, we are witnessing evidence for an ongoing tidal disruption at a rather early stage of its evolution. Our models predict that the ring should fill out at timescales proportional to the orbit, and since the orbit is not exceedingly long in this case, we might even predict that the entire filling of the disc could be monitored within a few decades.

For modeling the ZTF J013906.17+524536.89 transits, time step limitations make full SPH or N-body simulations such as that of \cite{VerasEtAl-2014} computationally difficult, if not impossible, certainly at high resolutions. With the hybrid model, however, we are not limited by the relatively large orbit, and these issues can be explored in depth in a future study.
 
\subsection{Producing the possible intact planet core of SDSS J1228+1040?}\label{SS:IntactCore}
The popular exoplanet-searching transit method has proven observationally difficult when used to search for planets around WDs, given how faint they typically are. Recently, \cite{ManserEtAl-2019} have used short-cadence spectroscopy in order to search for signs of random variations in the gaseous emission of a WD (SDSS J122859.93+104032) with a known gaseous disc, which they hypothesized would be produced by random gas-generating collisions among particles of dust. Instead they found periodic variations, which led the team to conclude that the emissions result from an excited cloud of gas trailing a metallic object up to a few hundred km in size, with a two-hour orbital period. Or else, the planetesimal itself is producing the gas since its orbit is close enough to the star (especially if it is eccentric), triggering surface sublimation. This discovery may turn out to become an additional important method for detecting planetary material. 

How might one produce a large intact iron object with the size of up to several hundered km, and an orbit with a pericentre distance ranging from $0.34R_{\odot}$ (if eccentric) to $0.73R_{\odot}$ (if circular), is an outstanding question \citep{ManserEtAl-2019}, and it is suggested that this object could be a left over planet core. Clearly, it must have reached that part of the disc \emph{after} the WD itself formed. The detailed formation simulation studies performed until now \citep{VerasEtAl-2014,VerasEtAl-2017} have considered neither the correct sizes, nor compositions. Our study has investigated for the first time the kind of parent objects that may indeed give rise to such planetary cores. Nevertheless, none of the planets that we have simulated has an intact surviving core. The mitigating circumstances could be that our simulated pericentre distances were insufficiently large to allow for the core survival by means of its higher density alone. More likely, we would require to repeat our hybrid simulations albeit utilizing SPH formulations that incorporate internal strength. Previous studies tentatively argue that an object held together by internal strength could indeed avoid disruption \citep{BearSoker-2013,BrownEtAl-2017,ManserEtAl-2019}. The hybrid model therefore has the potential to identify the most likely parent-objects leading to the \emph{initial} formation of a comparable iron core and its accompanying disc, which would later evolve to give rise to the SDSS J122859.93+104032 observables.

\section{Summary}\label{S:Summary}
In this study we employed a novel hybrid model introduced in an accompanying paper (Paper I) to study the typical debris discs that emerge from tidal disruptions of rocky dwarf and terrestrial planets, covering a range of masses, pericentre distances and semi-major axes between 3 AU and 150 AU. For the scenarios mentioned above, the resulting semi-major axis distribution at the end of disc formation bears a close resemblance to a \emph{bi-modal disruption} distribution, produced after the initial flyby of the disrupted progenitor. However, the departure from the analytical impulse approximation distribution is apparent when the pericentre distance is either a small (<10\%) fraction of the Roche limit or alternatively close to the Roche limit, which could be explained by major departures from the aforementioned impulse approximation assumptions. Our innermost orbit for the disc semi-major axis dispersion is approximately constant (to a factor of a 2), validating similar results for stellar tidal disruptions that also refuted the 'frozen-in' approximation in the energy spread of bound debris \citep{StoneEtAl-2013,GuillochonRamirez-Ruiz-2013,SteinbergEtAl-2019}.

We also show that repeated disruptions in the bound debris disc eventually flatten out the power-law size distribution of each individual disruption, finally settling on the numerical minimum of a single SPH particle. The timescale of disc formation completion in our hybrid models is compatible with that found in our full SPH simulations in Paper I. All of our hybrid simulations are 91-100 \% complete in just five orbital periods of the original progenitor. Tidal-spin up is also consistent with the full SPH simulations, although here we statistically quantify the phenomena. As much as 70-80\% of the fragments have rotation periods less than twice the known 2.2 h cohesionless asteroid spin-barrier. Resolution convergence in semi-major axis distribution is deemed relatively unimportant between 50K-500K SPH particles. 

Finally, the unbound mass fraction is around 50\% for all the disruptions modelled here. We characterize the properties of such ejected debris that then become interstellar asteroids, possibly similar to the recently observed 'Oumuammua' and 'Borisov' interstellar objects coming through the Solar system. Unbound fragments follow a power law size distribution with a $\sim$4/3 exponent, and we show that iron core fragments rotate faster than rocky fragments. We also find that tidal disruption of massive Earth-mass objects can eject very high velocity (up to several times the Sun's relative velocity) interstellar objects, which might provide a smoking gun signature for their origin.

Finally, we briefly discuss how our results, and in particular the bi-modal regime of tidal disruptions which might help explain the origins of Tabby's star variability, the transits of ZTF J0139+5245 and possible surviving planetary core around SDSS J122859+1040 from disruptions of dwarf-planets.

\section{Acknowledgment}\label{S:Acknowledgment}
We wish to thank the anonymous reviewer for excellent suggestions and comments that have greatly improved this manuscript. UM and HBP acknowledge support from the Minerva center for life under extreme planetary conditions, the Israeli Science and Technology ministry Ilan Ramon grant and the ISF I-CORE grant 1829/12.

\newpage
	
	
\bibliographystyle{mnras} 
\bibliography{bibfile}     

\begin{thebibliography}{}
\makeatletter
\relax
\def\mn@urlcharsother{\let\do\@makeother \do\$\do\&\do\#\do\^\do\_\do\%\do\~}
\def\mn@doi{\begingroup\mn@urlcharsother \@ifnextchar [ {\mn@doi@}
  {\mn@doi@[]}}
\def\mn@doi@[#1]#2{\def\@tempa{#1}\ifx\@tempa\@empty \href
  {http://dx.doi.org/#2} {doi:#2}\else \href {http://dx.doi.org/#2} {#1}\fi
  \endgroup}
\def\mn@eprint#1#2{\mn@eprint@#1:#2::\@nil}
\def\mn@eprint@arXiv#1{\href {http://arxiv.org/abs/#1} {{\tt arXiv:#1}}}
\def\mn@eprint@dblp#1{\href {http://dblp.uni-trier.de/rec/bibtex/#1.xml}
  {dblp:#1}}
\def\mn@eprint@#1:#2:#3:#4\@nil{\def\@tempa {#1}\def\@tempb {#2}\def\@tempc
  {#3}\ifx \@tempc \@empty \let \@tempc \@tempb \let \@tempb \@tempa \fi \ifx
  \@tempb \@empty \def\@tempb {arXiv}\fi \@ifundefined
  {mn@eprint@\@tempb}{\@tempb:\@tempc}{\expandafter \expandafter \csname
  mn@eprint@\@tempb\endcsname \expandafter{\@tempc}}}

\bibitem[\protect\citeauthoryear{{Bear} \& {Soker}}{{Bear} \&
  {Soker}}{2013}]{BearSoker-2013}
{Bear} E.,  {Soker} N.,  2013, \mn@doi [New Astronomy]
  {10.1016/j.newast.2012.08.004}, \href
  {http://adsabs.harvard.edu/abs/2013NewA...19...56B} {19, 56}

\bibitem[\protect\citeauthoryear{{Bergfors}, {Farihi}, {Dufour}  \&
  {Rocchetto}}{{Bergfors} et~al.}{2014}]{BergforsEtAl-2014}
{Bergfors} C.,  {Farihi} J.,  {Dufour} P.,   {Rocchetto} M.,  2014, \mn@doi
  [Monthly Notices of the Royal Astronomical Society] {10.1093/mnras/stu1565},
  \href {http://adsabs.harvard.edu/abs/2014MNRAS.444.2147B} {444, 2147}

\bibitem[\protect\citeauthoryear{{Bodman} \& {Quillen}}{{Bodman} \&
  {Quillen}}{2016}]{BodmanQuillen-2016}
{Bodman} E. H.~L.,  {Quillen} A.,  2016, \mn@doi [The Astrophysical Journal
  Letters] {10.3847/2041-8205/819/2/L34}, \href
  {https://ui.adsabs.harvard.edu/abs/2016ApJ...819L..34B} {819, L34}

\bibitem[\protect\citeauthoryear{{Bonsor}, {Mustill}  \& {Wyatt}}{{Bonsor}
  et~al.}{2011}]{BonsorEtAl-2011}
{Bonsor} A.,  {Mustill} A.~J.,   {Wyatt} M.~C.,  2011, \mn@doi [Monthly Notices
  of the Royal Astronomical Society] {10.1111/j.1365-2966.2011.18524.x}, \href
  {http://adsabs.harvard.edu/abs/2011MNRAS.414..930B} {414, 930}

\bibitem[\protect\citeauthoryear{{Boyajian} et~al.,}{{Boyajian}
  et~al.}{2016}]{BoyajianEtAl-2016}
{Boyajian} T.~S.,  et~al., 2016, \mn@doi [Monthly Notices of the Royal
  Astronomical Society] {10.1093/mnras/stw218}, \href
  {https://ui.adsabs.harvard.edu/abs/2016MNRAS.457.3988B} {457, 3988}

\bibitem[\protect\citeauthoryear{{Brown}, {Veras}  \& {G{\"a}nsicke}}{{Brown}
  et~al.}{2017}]{BrownEtAl-2017}
{Brown} J.~C.,  {Veras} D.,   {G{\"a}nsicke} B.~T.,  2017, \mn@doi [Monthly
  Notices of the Royal Astronomical Society] {10.1093/mnras/stx428}, \href
  {http://adsabs.harvard.edu/abs/2017MNRAS.468.1575B} {468, 1575}

\bibitem[\protect\citeauthoryear{{Burger}, {Maindl}  \& {Sch{\"a}fer}}{{Burger}
  et~al.}{2018}]{BurgerEtAl-2018}
{Burger} C.,  {Maindl} T.~I.,   {Sch{\"a}fer} C.~M.,  2018, \mn@doi [Celestial
  Mechanics and Dynamical Astronomy] {10.1007/s10569-017-9795-3}, \href
  {http://adsabs.harvard.edu/abs/2018CeMDA.130....2B} {130}

\bibitem[\protect\citeauthoryear{{Caiazzo} \& {Heyl}}{{Caiazzo} \&
  {Heyl}}{2017}]{CaiazzoHeyl-2017}
{Caiazzo} I.,  {Heyl} J.~S.,  2017, \mn@doi [Monthly Notices of the Royal
  Astronomical Society] {10.1093/mnras/stx1036}, \href
  {http://adsabs.harvard.edu/abs/2017MNRAS.469.2750C} {469, 2750}

\bibitem[\protect\citeauthoryear{{Debes} \& {Sigurdsson}}{{Debes} \&
  {Sigurdsson}}{2002}]{DebesSigurdsson-2002}
{Debes} J.~H.,  {Sigurdsson} S.,  2002, \mn@doi [The Astrophysical Journal]
  {10.1086/340291}, \href {http://adsabs.harvard.edu/abs/2002ApJ...572..556D}
  {572, 556}

\bibitem[\protect\citeauthoryear{{Debes}, {Walsh}  \& {Stark}}{{Debes}
  et~al.}{2012}]{DebesEtAl-2012}
{Debes} J.~H.,  {Walsh} K.~J.,   {Stark} C.,  2012, \mn@doi [The Astrophysical
  Journal] {10.1088/0004-637X/747/2/148}, \href
  {http://adsabs.harvard.edu/abs/2012ApJ...747..148D} {747, 148}

\bibitem[\protect\citeauthoryear{{Dennihy}, {Clemens}, {Dunlap}, {Fanale},
  {Fuchs}  \& {Hermes}}{{Dennihy} et~al.}{2018}]{DennihyEtAl-2018}
{Dennihy} E.,  {Clemens} J.~C.,  {Dunlap} B.~H.,  {Fanale} S.~M.,  {Fuchs}
  J.~T.,   {Hermes} J.~J.,  2018, \mn@doi [The Astrophysical Journal]
  {10.3847/1538-4357/aaa89b}, \href
  {https://ui.adsabs.harvard.edu/abs/2018ApJ...854...40D} {854, 40}

\bibitem[\protect\citeauthoryear{{Desharnais}, {Wesemael}, {Chayer}, {Kruk}  \&
  {Saffer}}{{Desharnais} et~al.}{2008}]{DesharnaisEtAl-2008}
{Desharnais} S.,  {Wesemael} F.,  {Chayer} P.,  {Kruk} J.~W.,   {Saffer} R.~A.,
   2008, \mn@doi [The Astrophysical Journal] {10.1086/523699}, \href
  {http://adsabs.harvard.edu/abs/2008ApJ...672..540D} {672, 540}

\bibitem[\protect\citeauthoryear{{Doyle}, {Young}, {Klein}, {Zuckerman}  \&
  {Schlichting}}{{Doyle} et~al.}{2019}]{DoyleEtAl-2019}
{Doyle} A.~E.,  {Young} E.~D.,  {Klein} B.,  {Zuckerman} B.,   {Schlichting}
  H.~E.,  2019, \mn@doi [Science] {10.1126/science.aax3901}, \href
  {https://ui.adsabs.harvard.edu/abs/2019Sci...366..356D} {366, 356}

\bibitem[\protect\citeauthoryear{{Dufour} et~al.,}{{Dufour}
  et~al.}{2007}]{DufourEtAl-2007}
{Dufour} P.,  et~al., 2007, \mn@doi [The Astrophysical Journal]
  {10.1086/518468}, \href {http://adsabs.harvard.edu/abs/2007ApJ...663.1291D}
  {663, 1291}

\bibitem[\protect\citeauthoryear{{Farihi}}{{Farihi}}{2016}]{Farihi-2016}
{Farihi} J.,  2016, \mn@doi [New Astronomy Reviews]
  {10.1016/j.newar.2016.03.001}, \href
  {http://adsabs.harvard.edu/abs/2016NewAR..71....9F} {71, 9}

\bibitem[\protect\citeauthoryear{{Fraser}, {Pravec}, {Fitzsimmons}, {Lacerda},
  {Bannister}, {Snodgrass}  \& {Smoli{\'c}}}{{Fraser}
  et~al.}{2018}]{FraserEtAl-2018}
{Fraser} W.~C.,  {Pravec} P.,  {Fitzsimmons} A.,  {Lacerda} P.,  {Bannister}
  M.~T.,  {Snodgrass} C.,   {Smoli{\'c}} I.,  2018, \mn@doi [Nature Astronomy]
  {10.1038/s41550-018-0398-z}, \href
  {https://ui.adsabs.harvard.edu/abs/2018NatAs...2..383F} {2, 383}

\bibitem[\protect\citeauthoryear{{G{\"a}nsicke}, {Koester}, {Farihi}, {Girven},
  {Parsons}  \& {Breedt}}{{G{\"a}nsicke} et~al.}{2012}]{GansickeEtAl-2012}
{G{\"a}nsicke} B.~T.,  {Koester} D.,  {Farihi} J.,  {Girven} J.,  {Parsons}
  S.~G.,   {Breedt} E.,  2012, \mn@doi [Monthly Notices of the Royal
  Astronomical Society] {10.1111/j.1365-2966.2012.21201.x}, \href
  {http://adsabs.harvard.edu/abs/2012MNRAS.424..333G} {424, 333}

\bibitem[\protect\citeauthoryear{{Grishin} \& {Veras}}{{Grishin} \&
  {Veras}}{2019}]{GrishinVeras-2019}
{Grishin} E.,  {Veras} D.,  2019, \mn@doi [Monthly Notices of the Royal
  Astronomical Society] {10.1093/mnras/stz2148}, \href
  {https://ui.adsabs.harvard.edu/abs/2019MNRAS.489..168G} {489, 168}

\bibitem[\protect\citeauthoryear{{Grishin}, {Perets}  \& {Avni}}{{Grishin}
  et~al.}{2019}]{GrishinEtAl-2019}
{Grishin} E.,  {Perets} H.~B.,   {Avni} Y.,  2019, \mn@doi [Monthly Notices of
  the Royal Astronomical Society] {10.1093/mnras/stz1505}, \href
  {https://ui.adsabs.harvard.edu/abs/2019MNRAS.487.3324G} {487, 3324}

\bibitem[\protect\citeauthoryear{{Guillochon} \& {Ramirez-Ruiz}}{{Guillochon}
  \& {Ramirez-Ruiz}}{2013}]{GuillochonRamirez-Ruiz-2013}
{Guillochon} J.,  {Ramirez-Ruiz} E.,  2013, \mn@doi [The Astrophysical Journal]
  {10.1088/0004-637X/767/1/25}, \href
  {https://ui.adsabs.harvard.edu/abs/2013ApJ...767...25G} {767, 25}

\bibitem[\protect\citeauthoryear{{Hamers} \& {Portegies Zwart}}{{Hamers} \&
  {Portegies Zwart}}{2016}]{HamersPortegiesZwart-2016}
{Hamers} A.~S.,  {Portegies Zwart} S.~F.,  2016, \mn@doi [Monthly Notices of
  the Royal Astronomical Society] {10.1093/mnrasl/slw134}, \href
  {http://adsabs.harvard.edu/abs/2016MNRAS.462L..84H} {462, L84}

\bibitem[\protect\citeauthoryear{{Harrison}, {Bonsor}  \&
  {Madhusudhan}}{{Harrison} et~al.}{2018}]{HarrisonEtAl-2018}
{Harrison} J. H.~D.,  {Bonsor} A.,   {Madhusudhan} N.,  2018, \mn@doi [Monthly
  Notices of the Royal Astronomical Society] {10.1093/mnras/sty1700}, \href
  {https://ui.adsabs.harvard.edu/abs/2018MNRAS.479.3814H} {479, 3814}

\bibitem[\protect\citeauthoryear{{Hollands}, {G{\"a}nsicke}  \&
  {Koester}}{{Hollands} et~al.}{2018}]{HollandsEtAl-2018}
{Hollands} M.~A.,  {G{\"a}nsicke} B.~T.,   {Koester} D.,  2018, \mn@doi
  [Monthly Notices of the Royal Astronomical Society] {10.1093/mnras/sty592},
  \href {https://ui.adsabs.harvard.edu/abs/2018MNRAS.477...93H} {477, 93}

\bibitem[\protect\citeauthoryear{{Jura}}{{Jura}}{2003}]{Jura-2003}
{Jura} M.,  2003, \mn@doi [The Astrophysical Journal] {10.1086/374036}, \href
  {http://adsabs.harvard.edu/abs/2003ApJ...584L..91J} {584, L91}

\bibitem[\protect\citeauthoryear{{Jura}}{{Jura}}{2008}]{Jura-2008}
{Jura} M.,  2008, \mn@doi [The Astronomical Journal]
  {10.1088/0004-6256/135/5/1785}, \href
  {http://adsabs.harvard.edu/abs/2008AJ....135.1785J} {135, 1785}

\bibitem[\protect\citeauthoryear{{Jura} \& {Young}}{{Jura} \&
  {Young}}{2014}]{JuraYoung-2014}
{Jura} M.,  {Young} E.~D.,  2014, \mn@doi [Annual Review of Earth and Planetary
  Sciences] {10.1146/annurev-earth-060313-054740}, \href
  {http://adsabs.harvard.edu/abs/2014AREPS..42...45J} {42, 45}

\bibitem[\protect\citeauthoryear{{Jura}, {Farihi}, {Zuckerman}  \&
  {Becklin}}{{Jura} et~al.}{2007}]{JuraEtAl-2007}
{Jura} M.,  {Farihi} J.,  {Zuckerman} B.,   {Becklin} E.~E.,  2007, \mn@doi
  [The Astronomical Journal] {10.1086/512734}, \href
  {http://adsabs.harvard.edu/abs/2007AJ....133.1927J} {133, 1927}

\bibitem[\protect\citeauthoryear{{Jura}, {Farihi}  \& {Zuckerman}}{{Jura}
  et~al.}{2009}]{JuraEtAl-2009}
{Jura} M.,  {Farihi} J.,   {Zuckerman} B.,  2009, \mn@doi [The Astronomical
  Journal] {10.1088/0004-6256/137/2/3191}, \href
  {http://adsabs.harvard.edu/abs/2009AJ....137.3191J} {137, 3191}

\bibitem[\protect\citeauthoryear{{Kilic}, {von Hippel}, {Leggett}  \&
  {Winget}}{{Kilic} et~al.}{2006}]{KilicEtAl-2006}
{Kilic} M.,  {von Hippel} T.,  {Leggett} S.~K.,   {Winget} D.~E.,  2006,
  \mn@doi [The Astrophysical Journal] {10.1086/504682}, \href
  {http://adsabs.harvard.edu/abs/2006ApJ...646..474K} {646, 474}

\bibitem[\protect\citeauthoryear{{Klein}, {Jura}, {Koester}, {Zuckerman}  \&
  {Melis}}{{Klein} et~al.}{2010}]{KleinEtAl-2010}
{Klein} B.,  {Jura} M.,  {Koester} D.,  {Zuckerman} B.,   {Melis} C.,  2010,
  \mn@doi [The Astrophysical Journal] {10.1088/0004-637X/709/2/950}, \href
  {http://adsabs.harvard.edu/abs/2010ApJ...709..950K} {709, 950}

\bibitem[\protect\citeauthoryear{{Koester}, {G{\"a}nsicke}  \&
  {Farihi}}{{Koester} et~al.}{2014}]{KoesterEtAl-2014}
{Koester} D.,  {G{\"a}nsicke} B.~T.,   {Farihi} J.,  2014, \mn@doi [Astronomy
  and Astrophysics] {10.1051/0004-6361/201423691}, \href
  {http://adsabs.harvard.edu/abs/2014A%26A...566A..34K} {566, A34}

\bibitem[\protect\citeauthoryear{{Kratter} \& {Perets}}{{Kratter} \&
  {Perets}}{2012}]{KratterPerets-2012}
{Kratter} K.~M.,  {Perets} H.~B.,  2012, \mn@doi [The Astrophysical Journal]
  {10.1088/0004-637X/753/1/91}, \href
  {http://adsabs.harvard.edu/abs/2012ApJ...753...91K} {753, 91}

\bibitem[\protect\citeauthoryear{{Makarov} \& {Veras}}{{Makarov} \&
  {Veras}}{2019}]{MakarovVeras-2019}
{Makarov} V.~V.,  {Veras} D.,  2019, arXiv e-prints, \href
  {https://ui.adsabs.harvard.edu/abs/2019arXiv190804612M} {p. arXiv:1908.04612}

\bibitem[\protect\citeauthoryear{{Malamud} \& {Prialnik}}{{Malamud} \&
  {Prialnik}}{2015}]{MalamudPrialnik-2015}
{Malamud} U.,  {Prialnik} D.,  2015, \mn@doi [Icarus]
  {10.1016/j.icarus.2014.02.027}, \href
  {http://adsabs.harvard.edu/abs/2015Icar..246...21M} {246, 21}

\bibitem[\protect\citeauthoryear{{Malamud}, {Perets}  \& {Schubert}}{{Malamud}
  et~al.}{2017}]{MalamudEtAl-2017}
{Malamud} U.,  {Perets} H.~B.,   {Schubert} G.,  2017, \mn@doi [Monthly Notices
  of the Royal Astronomical Society] {10.1093/mnras/stx546}, \href
  {http://adsabs.harvard.edu/abs/2017MNRAS.468.1056M} {468, 1056}

\bibitem[\protect\citeauthoryear{{Malamud}, {Perets}, {Sch{\"a}fer}  \&
  {Burger}}{{Malamud} et~al.}{2018}]{MalamudEtAl-2018}
{Malamud} U.,  {Perets} H.~B.,  {Sch{\"a}fer} C.,   {Burger} C.,  2018, \mn@doi
  [Monthly Notices of the Royal Astronomical Society] {10.1093/mnras/sty1667},
  \href {http://adsabs.harvard.edu/abs/2018MNRAS.479.1711M} {479, 1711}

\bibitem[\protect\citeauthoryear{{Manser} et~al.,}{{Manser}
  et~al.}{2016}]{ManserEtAl-2016}
{Manser} C.~J.,  et~al., 2016, \mn@doi [Monthly Notices of the Royal
  Astronomical Society] {10.1093/mnras/stv2603}, \href
  {http://adsabs.harvard.edu/abs/2016MNRAS.455.4467M} {455, 4467}

\bibitem[\protect\citeauthoryear{{Manser} et~al.,}{{Manser}
  et~al.}{2019}]{ManserEtAl-2019}
{Manser} C.~J.,  et~al., 2019, \mn@doi [Science] {10.1126/science.aat5330},
  \href {https://ui.adsabs.harvard.edu/abs/2019Sci...364...66M} {364, 66}

\bibitem[\protect\citeauthoryear{{Meech} et~al.,}{{Meech}
  et~al.}{2017}]{MeechEtAl-2017}
{Meech} K.~J.,  et~al., 2017, \mn@doi [Nature] {10.1038/nature25020}, \href
  {https://ui.adsabs.harvard.edu/abs/2017Natur.552..378M} {552, 378}

\bibitem[\protect\citeauthoryear{{Metzger}, {Rafikov}  \&
  {Bochkarev}}{{Metzger} et~al.}{2012}]{MetzgerEtAl-2012}
{Metzger} B.~D.,  {Rafikov} R.~R.,   {Bochkarev} K.~V.,  2012, \mn@doi [Monthly
  Notices of the Royal Astronomical Society]
  {10.1111/j.1365-2966.2012.20895.x}, \href
  {http://adsabs.harvard.edu/abs/2012MNRAS.423..505M} {423, 505}

\bibitem[\protect\citeauthoryear{{Michaely} \& {Perets}}{{Michaely} \&
  {Perets}}{2014}]{MichaelyPerets-2014}
{Michaely} E.,  {Perets} H.~B.,  2014, \mn@doi [The Astrophysical Journal]
  {10.1088/0004-637X/794/2/122}, \href
  {http://adsabs.harvard.edu/abs/2014ApJ...794..122M} {794, 122}

\bibitem[\protect\citeauthoryear{{Montet} \& {Simon}}{{Montet} \&
  {Simon}}{2016}]{MontetSimon-2016}
{Montet} B.~T.,  {Simon} J.~D.,  2016, \mn@doi [The Astrophysical Journal
  Letters] {10.3847/2041-8205/830/2/L39}, \href
  {https://ui.adsabs.harvard.edu/abs/2016ApJ...830L..39M} {830, L39}

\bibitem[\protect\citeauthoryear{{Mustill} \& {Villaver}}{{Mustill} \&
  {Villaver}}{2012}]{MustillVillaver-2012}
{Mustill} A.~J.,  {Villaver} E.,  2012, \mn@doi [The Astrophysical Journal]
  {10.1088/0004-637X/761/2/121}, \href
  {http://adsabs.harvard.edu/abs/2012ApJ...761..121M} {761, 121}

\bibitem[\protect\citeauthoryear{{Neslu{\v{s}}an} \& {Budaj}}{{Neslu{\v{s}}an}
  \& {Budaj}}{2017}]{NeslusanBudaj-2017}
{Neslu{\v{s}}an} L.,  {Budaj} J.,  2017, \mn@doi [Astronomy \& Astrophysics]
  {10.1051/0004-6361/201629344}, \href
  {https://ui.adsabs.harvard.edu/abs/2017A&A...600A..86N} {600, A86}

\bibitem[\protect\citeauthoryear{{Payne}, {Veras}, {Holman}  \&
  {G{\"a}nsicke}}{{Payne} et~al.}{2016}]{PayneEtAl-2016}
{Payne} M.~J.,  {Veras} D.,  {Holman} M.~J.,   {G{\"a}nsicke} B.~T.,  2016,
  \mn@doi [Monthly Notices of the Royal Astronomical Society]
  {10.1093/mnras/stv2966}, \href
  {http://adsabs.harvard.edu/abs/2016MNRAS.457..217P} {457, 217}

\bibitem[\protect\citeauthoryear{{Payne}, {Veras}, {G{\"a}nsicke}  \&
  {Holman}}{{Payne} et~al.}{2017}]{PayneEtAl-2017}
{Payne} M.~J.,  {Veras} D.,  {G{\"a}nsicke} B.~T.,   {Holman} M.~J.,  2017,
  \mn@doi [Monthly Notices of the Royal Astronomical Society]
  {10.1093/mnras/stw2585}, \href
  {http://adsabs.harvard.edu/abs/2017MNRAS.464.2557P} {464, 2557}

\bibitem[\protect\citeauthoryear{{Perets} \& {Gualandris}}{{Perets} \&
  {Gualandris}}{2010}]{PeretsGualandris-2010}
{Perets} H.~B.,  {Gualandris} A.,  2010, \mn@doi [The Astrophysical Journal]
  {10.1088/0004-637X/719/1/220}, \href
  {http://adsabs.harvard.edu/abs/2010ApJ...719..220P} {719, 220}

\bibitem[\protect\citeauthoryear{{Perets} \& {Kratter}}{{Perets} \&
  {Kratter}}{2012}]{PeretsKratter-2012}
{Perets} H.~B.,  {Kratter} K.~M.,  2012, \mn@doi [The Astrophysical Journal]
  {10.1088/0004-637X/760/2/99}, \href
  {http://adsabs.harvard.edu/abs/2012ApJ...760...99P} {760, 99}

\bibitem[\protect\citeauthoryear{{Petrovich} \& {Mu{\~n}oz}}{{Petrovich} \&
  {Mu{\~n}oz}}{2017}]{PetrovichMunoz-2017}
{Petrovich} C.,  {Mu{\~n}oz} D.~J.,  2017, \mn@doi [Astrophysical Journal]
  {10.3847/1538-4357/834/2/116}, \href
  {http://adsabs.harvard.edu/abs/2017ApJ...834..116P} {834, 116}

\bibitem[\protect\citeauthoryear{{Pravec}, {Harris}  \& {Michalowski}}{{Pravec}
  et~al.}{2002}]{PravecEtAl-2002}
{Pravec} P.,  {Harris} A.~W.,   {Michalowski} T.,  2002, {Asteroid Rotations}.
pp 113--122

\bibitem[\protect\citeauthoryear{{Rafikov}}{{Rafikov}}{2011}]{Rafikov-2011}
{Rafikov} R.~R.,  2011, \mn@doi [Monthly Notices of the Royal Astronomical
  Society: Letters] {10.1111/j.1745-3933.2011.01096.x}, \href
  {http://adsabs.harvard.edu/abs/2011MNRAS.416L..55R} {416, L55}

\bibitem[\protect\citeauthoryear{{Reach}, {Kuchner}, {von Hippel}, {Burrows},
  {Mullally}, {Kilic}  \& {Winget}}{{Reach} et~al.}{2005}]{ReachEtAl-2005}
{Reach} W.~T.,  {Kuchner} M.~J.,  {von Hippel} T.,  {Burrows} A.,  {Mullally}
  F.,  {Kilic} M.,   {Winget} D.~E.,  2005, \mn@doi [The Astrophysical Journal]
  {10.1086/499561}, \href {http://adsabs.harvard.edu/abs/2005ApJ...635L.161R}
  {635, L161}

\bibitem[\protect\citeauthoryear{{Reach}, {Lisse}, {von Hippel}  \&
  {Mullally}}{{Reach} et~al.}{2009}]{ReachEtAl-2009}
{Reach} W.~T.,  {Lisse} C.,  {von Hippel} T.,   {Mullally} F.,  2009, \mn@doi
  [The Astrophysical Journal] {10.1088/0004-637X/693/1/697}, \href
  {http://adsabs.harvard.edu/abs/2009ApJ...693..697R} {693, 697}

\bibitem[\protect\citeauthoryear{{Rickman}, {Fouchard}, {Froeschl{\'e}}  \&
  {Valsecchi}}{{Rickman} et~al.}{2008}]{RickmanEtAl-2008}
{Rickman} H.,  {Fouchard} M.,  {Froeschl{\'e}} C.,   {Valsecchi} G.~B.,  2008,
  \mn@doi [Celestial Mechanics and Dynamical Astronomy]
  {10.1007/s10569-008-9140-y}, \href
  {http://adsabs.harvard.edu/abs/2008CeMDA.102..111R} {102, 111}

\bibitem[\protect\citeauthoryear{{Sacco}, {Ngo}  \& {Modolo}}{{Sacco}
  et~al.}{2018}]{SaccoEtAl-2018}
{Sacco} G.,  {Ngo} L.~D.,   {Modolo} J.,  2018, Journal of the American
  Association of Variable Star Observers (JAAVSO), \href
  {https://ui.adsabs.harvard.edu/abs/2018JAVSO..46...14S} {46, 14}

\bibitem[\protect\citeauthoryear{{Sari}, {Kobayashi}  \& {Rossi}}{{Sari}
  et~al.}{2010}]{SariEtAl-2010}
{Sari} R.,  {Kobayashi} S.,   {Rossi} E.~M.,  2010, \mn@doi [The Astrophysical
  Journal] {10.1088/0004-637X/708/1/605}, \href
  {https://ui.adsabs.harvard.edu/abs/2010ApJ...708..605S} {708, 605}

\bibitem[\protect\citeauthoryear{{Schaefer}}{{Schaefer}}{2016}]{Schaefer-2016}
{Schaefer} B.~E.,  2016, \mn@doi [The Astrophysical Journal Letters]
  {10.3847/2041-8205/822/2/L34}, \href
  {https://ui.adsabs.harvard.edu/abs/2016ApJ...822L..34S} {822, L34}

\bibitem[\protect\citeauthoryear{{Schaefer}}{{Schaefer}}{2019}]{Schaefer-2019}
{Schaefer} B.~E.,  2019, \mn@doi [Research Notes of the American Astronomical
  Society] {10.3847/2515-5172/ab21dd}, \href
  {https://ui.adsabs.harvard.edu/abs/2019RNAAS...3...77S} {3, 77}

\bibitem[\protect\citeauthoryear{{Sch{\"a}fer}, {Riecker}, {Maindl}, {Speith},
  {Scherrer}  \& {Kley}}{{Sch{\"a}fer} et~al.}{2016}]{SchaferEtAl-2016}
{Sch{\"a}fer} C.,  {Riecker} S.,  {Maindl} T.~I.,  {Speith} R.,  {Scherrer} S.,
    {Kley} W.,  2016, \mn@doi [Astronomy \& Astrophysics]
  {10.1051/0004-6361/201528060}, \href
  {http://adsabs.harvard.edu/abs/2016A%26A...590A..19S} {590, A19}

\bibitem[\protect\citeauthoryear{{Scheeres}, {Ostro}, {Werner}, {Asphaug}  \&
  {Hudson}}{{Scheeres} et~al.}{2000}]{ScheeresEtAl-2000}
{Scheeres} D.~J.,  {Ostro} S.~J.,  {Werner} R.~A.,  {Asphaug} E.,   {Hudson}
  R.~S.,  2000, \mn@doi [Icarus] {10.1006/icar.2000.6443}, \href
  {https://ui.adsabs.harvard.edu/abs/2000Icar..147..106S} {147, 106}

\bibitem[\protect\citeauthoryear{{Shappee} \& {Thompson}}{{Shappee} \&
  {Thompson}}{2013}]{ShapeeThompson-2013}
{Shappee} B.~J.,  {Thompson} T.~A.,  2013, \mn@doi [The Astrophysical Journal]
  {10.1088/0004-637X/766/1/64}, \href
  {http://adsabs.harvard.edu/abs/2013ApJ...766...64S} {766, 64}

\bibitem[\protect\citeauthoryear{{Siraj} \& {Loeb}}{{Siraj} \&
  {Loeb}}{2019}]{SirajLoeb-2019}
{Siraj} A.,  {Loeb} A.,  2019, arXiv e-prints, \href
  {https://ui.adsabs.harvard.edu/abs/2019arXiv190407224S} {p. arXiv:1904.07224}

\bibitem[\protect\citeauthoryear{{Smallwood}, {Martin}, {Livio}  \&
  {Lubow}}{{Smallwood} et~al.}{2018}]{SmallwoodEtAl-2018}
{Smallwood} J.~L.,  {Martin} R.~G.,  {Livio} M.,   {Lubow} S.~H.,  2018,
  \mn@doi [Monthly Notices of the Royal Astronomical Society]
  {10.1093/mnras/sty1819}, \href
  {https://ui.adsabs.harvard.edu/abs/2018MNRAS.480...57S} {480, 57}

\bibitem[\protect\citeauthoryear{{Steinberg}, {Coughlin}, {Stone}  \&
  {Metzger}}{{Steinberg} et~al.}{2019}]{SteinbergEtAl-2019}
{Steinberg} E.,  {Coughlin} E.~R.,  {Stone} N.~C.,   {Metzger} B.~D.,  2019,
  \mn@doi [Monthly Notices of the Royal Astronomical Society]
  {10.1093/mnrasl/slz048}, \href
  {https://ui.adsabs.harvard.edu/abs/2019MNRAS.485L.146S} {485, L146}

\bibitem[\protect\citeauthoryear{{Stephan}, {Naoz}  \& {Zuckerman}}{{Stephan}
  et~al.}{2017}]{StephanEtAl-2017}
{Stephan} A.~P.,  {Naoz} S.,   {Zuckerman} B.,  2017, \mn@doi [The
  Astrophysical Journal Letters] {10.3847/2041-8213/aa7cf3}, \href
  {http://adsabs.harvard.edu/abs/2017ApJ...844L..16S} {844, L16}

\bibitem[\protect\citeauthoryear{{Stone}, {Sari}  \& {Loeb}}{{Stone}
  et~al.}{2013}]{StoneEtAl-2013}
{Stone} N.,  {Sari} R.,   {Loeb} A.,  2013, \mn@doi [Monthly Notices of the
  Royal Astronomical Society] {10.1093/mnras/stt1270}, \href
  {https://ui.adsabs.harvard.edu/abs/2013MNRAS.435.1809S} {435, 1809}

\bibitem[\protect\citeauthoryear{{Stone}, {Metzger}  \& {Loeb}}{{Stone}
  et~al.}{2015}]{StoneEtAl-2015}
{Stone} N.,  {Metzger} B.~D.,   {Loeb} A.,  2015, \mn@doi [Monthly Notices of
  the Royal Astronomical Society] {10.1093/mnras/stu2718}, \href
  {http://adsabs.harvard.edu/abs/2015MNRAS.448..188S} {448, 188}

\bibitem[\protect\citeauthoryear{{Swan}, {Farihi}, {Koester}, {Holland s},
  {Parsons}, {Cauley}, {Redfield}  \& {G{\"a}nsicke}}{{Swan}
  et~al.}{2019}]{SwanEtAl-2019}
{Swan} A.,  {Farihi} J.,  {Koester} D.,  {Holland s} M.,  {Parsons} S.,
  {Cauley} P.~W.,  {Redfield} S.,   {G{\"a}nsicke} B.~T.,  2019, \mn@doi
  [Monthly Notices of the Royal Astronomical Society] {10.1093/mnras/stz2337},
  \href {https://ui.adsabs.harvard.edu/abs/2019MNRAS.490..202S} {490, 202}

\bibitem[\protect\citeauthoryear{{Thompson} et~al.,}{{Thompson}
  et~al.}{2016}]{ThompsonEtAl-2016}
{Thompson} M.~A.,  et~al., 2016, \mn@doi [Monthly Notices of the Royal
  Astronomical Society] {10.1093/mnrasl/slw008}, \href
  {https://ui.adsabs.harvard.edu/abs/2016MNRAS.458L..39T} {458, L39}

\bibitem[\protect\citeauthoryear{{Vanderbosch} et~al.,}{{Vanderbosch}
  et~al.}{2019}]{VanderboschEtAl-2019}
{Vanderbosch} Z.,  et~al., 2019, arXiv e-prints, \href
  {https://ui.adsabs.harvard.edu/abs/2019arXiv190809839V} {p. arXiv:1908.09839}

\bibitem[\protect\citeauthoryear{{Vanderburg} et~al.,}{{Vanderburg}
  et~al.}{2015}]{VanderburgEtAl-2015}
{Vanderburg} A.,  et~al., 2015, \mn@doi [Nature] {10.1038/nature15527}, \href
  {http://adsabs.harvard.edu/abs/2015Natur.526..546V} {526, 546}

\bibitem[\protect\citeauthoryear{{Veras}}{{Veras}}{2016}]{Veras-2016}
{Veras} D.,  2016, \mn@doi [Royal Society Open Science] {10.1098/rsos.150571},
  \href {http://adsabs.harvard.edu/abs/2016RSOS....3.0571V} {3, 150571}

\bibitem[\protect\citeauthoryear{{Veras} \& {G{\"a}nsicke}}{{Veras} \&
  {G{\"a}nsicke}}{2015}]{VerasGansicke-2015}
{Veras} D.,  {G{\"a}nsicke} B.~T.,  2015, \mn@doi [Monthly Notices of the Royal
  Astronomical Society] {10.1093/mnras/stu2475}, \href
  {http://adsabs.harvard.edu/abs/2015MNRAS.447.1049V} {447, 1049}

\bibitem[\protect\citeauthoryear{{Veras}, {Leinhardt}, {Bonsor}  \&
  {G{\"a}nsicke}}{{Veras} et~al.}{2014}]{VerasEtAl-2014}
{Veras} D.,  {Leinhardt} Z.~M.,  {Bonsor} A.,   {G{\"a}nsicke} B.~T.,  2014,
  \mn@doi [Monthly Notices of the Royal Astronomical Society]
  {10.1093/mnras/stu1871}, \href
  {http://adsabs.harvard.edu/abs/2014MNRAS.445.2244V} {445, 2244}

\bibitem[\protect\citeauthoryear{{Veras}, {Leinhardt}, {Eggl}  \&
  {G{\"a}nsicke}}{{Veras} et~al.}{2015}]{VerasEtAl-2015}
{Veras} D.,  {Leinhardt} Z.~M.,  {Eggl} S.,   {G{\"a}nsicke} B.~T.,  2015,
  \mn@doi [Monthly Notices of the Royal Astronomical Society]
  {10.1093/mnras/stv1195}, \href
  {http://adsabs.harvard.edu/abs/2015MNRAS.451.3453V} {451, 3453}

\bibitem[\protect\citeauthoryear{{Veras}, {Carter}, {Leinhardt}  \&
  {G{\"a}nsicke}}{{Veras} et~al.}{2017}]{VerasEtAl-2017}
{Veras} D.,  {Carter} P.~J.,  {Leinhardt} Z.~M.,   {G{\"a}nsicke} B.~T.,  2017,
  \mn@doi [Monthly Notices of the Royal Astronomical Society]
  {10.1093/mnras/stw2748}, \href
  {https://ui.adsabs.harvard.edu/abs/2017MNRAS.465.1008V} {465, 1008}

\bibitem[\protect\citeauthoryear{{Villaver}, {Livio}, {Mustill}  \&
  {Siess}}{{Villaver} et~al.}{2014}]{VillaverEtAl-2014}
{Villaver} E.,  {Livio} M.,  {Mustill} A.~J.,   {Siess} L.,  2014, \mn@doi [The
  Astrophysical Journal] {10.1088/0004-637X/794/1/3}, \href
  {http://adsabs.harvard.edu/abs/2014ApJ...794....3V} {794, 3}

\bibitem[\protect\citeauthoryear{{Wolff}, {Koester}  \& {Liebert}}{{Wolff}
  et~al.}{2002}]{WolffEtAl-2002}
{Wolff} B.,  {Koester} D.,   {Liebert} J.,  2002, \mn@doi [Astronomy and
  Astrophysics] {10.1051/0004-6361:20020194}, \href
  {http://adsabs.harvard.edu/abs/2002A%26A...385..995W} {385, 995}

\bibitem[\protect\citeauthoryear{{Wyatt}, {van Lieshout}, {Kennedy}  \&
  {Boyajian}}{{Wyatt} et~al.}{2018}]{WyattEtAl-2018}
{Wyatt} M.~C.,  {van Lieshout} R.,  {Kennedy} G.~M.,   {Boyajian} T.~S.,  2018,
  \mn@doi [Monthly Notices of the Royal Astronomical Society]
  {10.1093/mnras/stx2713}, \href
  {https://ui.adsabs.harvard.edu/abs/2018MNRAS.473.5286W} {473, 5286}

\bibitem[\protect\citeauthoryear{{Yu} \& {Tremaine}}{{Yu} \&
  {Tremaine}}{2003}]{YuTremaine-2003}
{Yu} Q.,  {Tremaine} S.,  2003, \mn@doi [The Astrophysical Journal]
  {10.1086/379546}, \href
  {https://ui.adsabs.harvard.edu/abs/2003ApJ...599.1129Y} {599, 1129}

\bibitem[\protect\citeauthoryear{{Zuckerman}, {Koester}, {Reid}  \&
  {H{\"u}nsch}}{{Zuckerman} et~al.}{2003}]{ZuckermanEtAl-2003}
{Zuckerman} B.,  {Koester} D.,  {Reid} I.~N.,   {H{\"u}nsch} M.,  2003, \mn@doi
  [The Astrophysical Journal,] {10.1086/377492}, \href
  {http://adsabs.harvard.edu/abs/2003ApJ...596..477Z} {596, 477}

\bibitem[\protect\citeauthoryear{{Zuckerman}, {Melis}, {Klein}, {Koester}  \&
  {Jura}}{{Zuckerman} et~al.}{2010}]{ZuckermanEtAl-2010}
{Zuckerman} B.,  {Melis} C.,  {Klein} B.,  {Koester} D.,   {Jura} M.,  2010,
  \mn@doi [The Astrophysical Journal] {10.1088/0004-637X/722/1/725}, \href
  {http://adsabs.harvard.edu/abs/2010ApJ...722..725Z} {722, 725}

\bibitem[\protect\citeauthoryear{{de Le{\'o}n}, {Licandro}, {Serra-Ricart},
  {Cabrera-Lavers}, {Font Serra}, {Scarpa}, {de la Fuente Marcos}  \& {de la
  Fuente Marcos}}{{de Le{\'o}n} et~al.}{2019}]{DeLeonEtAl-2019}
{de Le{\'o}n} J.,  {Licandro} J.,  {Serra-Ricart} M.,  {Cabrera-Lavers} A.,
  {Font Serra} J.,  {Scarpa} R.,  {de la Fuente Marcos} C.,   {de la Fuente
  Marcos} R.,  2019, \mn@doi [Research Notes of the American Astronomical
  Society] {10.3847/2515-5172/ab449c}, \href
  {https://ui.adsabs.harvard.edu/abs/2019RNAAS...3..131D} {3, 131}

\makeatother
\end{thebibliography}
\begin{appendices}	
	
\section{Resolution effects on disc size distribution}\label{A:ResolutionDiscSizeDist}
\begin{figure}
 	\begin{center}	
 		\subfigure{\includegraphics[scale=0.53]{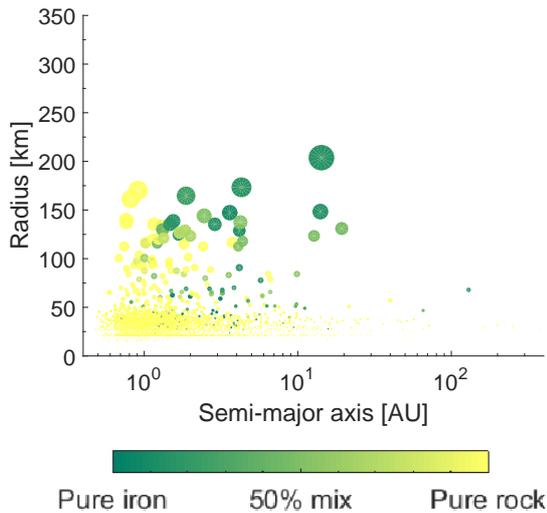}}
 		\subfigure{\includegraphics[scale=0.53]{ResLegend.eps}}
 		\caption{Fragment size distribution after the complete disc formation using the hybrid model, for a tidally disrupted Kuiper-belt dwarf-planet analogue ($a=150$ AU and $q=1R_{\odot}$) around a 0.6$M_{\odot}$ WD. Fragment radii are plotted as a function of their semi-major axis. Each fragment is depicted by a circle whose size is scaled with the number of its constituent SPH particles, and the colour coding represents different compositions (see underlying colorbar).}
 		\label{fig:150AUSizeDist} 		
 	\end{center}	
 \end{figure}

 In the main text we discuss the gradual flatenning of the bound disc fragment size distribution. All of our simulations for $q=0.1R_{\odot}$ and $q=0.5R_{\odot}$ have indeed shown a similar pattern of behaviour, which over time reduces the size of tidally disrupted particles down to the numerical minimum. This sequence proceeds gradually from the inside out. Each disruption generates in itself a size distribution which roughly follows a power law behavior, but after many subsequent disruptions the distribution becomes flat. For $q=1R_{\odot}$, the outcome of the hybrid model is however not identical to those with $q=0.5R_{\odot}$ and $q=0.1R_{\odot}$. Even partial disruptions should, eventually, strip material bit by bit until reaching the minimum size. In Figure \ref{fig:150AUSizeDist}, however, we show an example with a different outcome. Here we simulate a Kuiper-belt analogue at $a=150$ AU and $q=1R_{\odot}$, using the hybrid model. All other models with $q=1R_{\odot}$ qualitatively result in the same outcome as Figure \ref{fig:150AUSizeDist}. The plot shows the size distribution of the fragments at the \emph{end} of the simulation. Fragments that are as large as 150-200 km and have relatively tighter orbits compared with the original object, are passing within the tidal sphere \emph{without} being disrupted, which eventually leads to the cessation of the simulation.

\begin{figure}
 	\begin{center}		
 		\subfigure[t=13000 s.] {\label{fig:1q0_75a13000s}\includegraphics[scale=0.235]{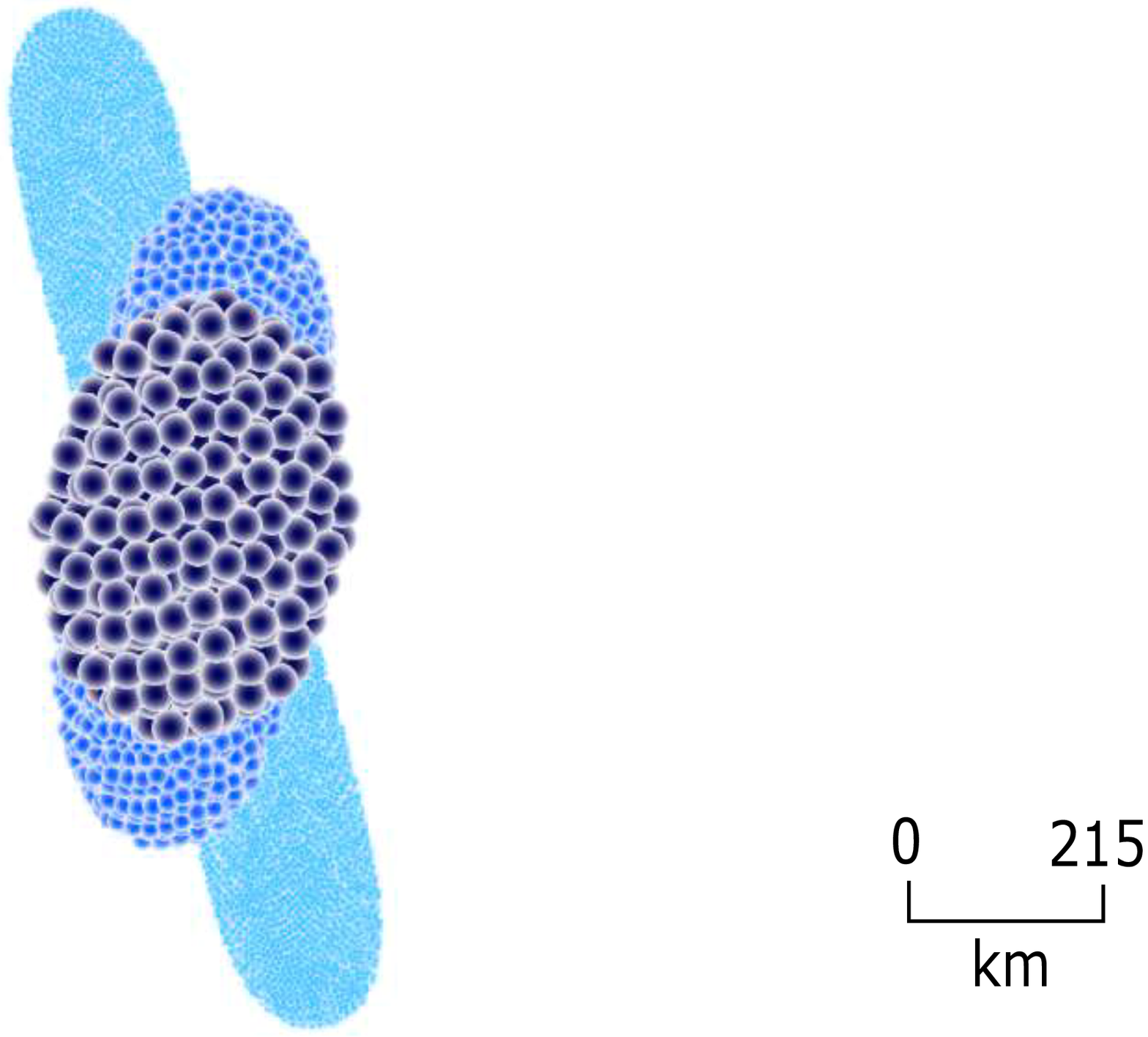}}
 		\subfigure[t=62000 s.] {\label{fig:1q0_75a62000s}\includegraphics[scale=0.235]{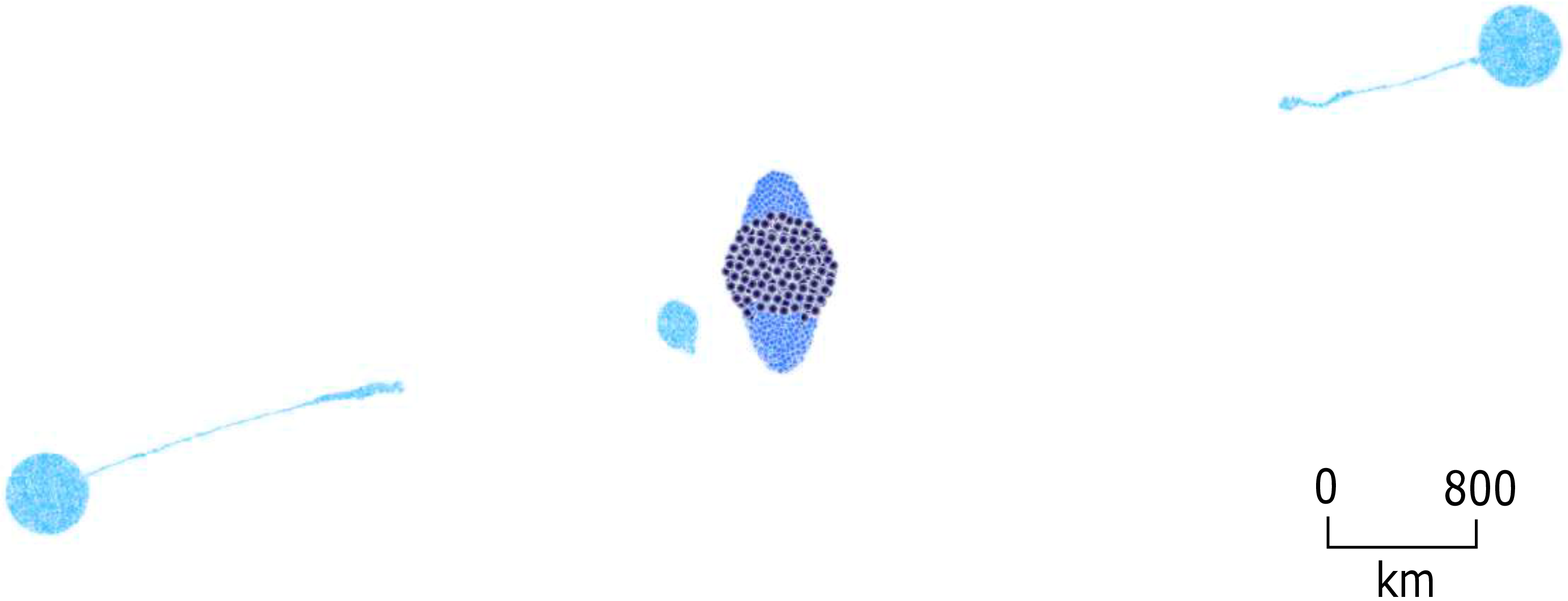}}
 	\end{center}
 	\caption{A top-down snapshot of partial tidal disruptions of three super imposed pure rocky fragments with $a=0.75$ AU and a pericentre distance of $q=1R_{\odot}$, around a 0.6$M_\odot$ WD. The fragments differ only in resolution: having 0.5K (dark-blue), 5K (blue) and 50K (light-blue) SPH particles respectively. The disruption outcome is shown (a) after the initial tidal encounter outside the Roche limit, and (b) after the fragmentation phase. Only the high resolution simulation results in the expected outcome of tidal breakup into 3 sub-fragments.}
 	\label{fig:1q0_75a}
 \end{figure}

 This result is puzzling, since we expect the disruptions to continue, and the distribution to flatten, as previously mentioned. Here we report that the reason for this outcome is numerical, and indeed not physical, as follows. Given the initial resolution of the hybrid simulation of 200K SPH particles, the number of SPH particles comprising even the largest of the fragments in Figure \ref{fig:150AUSizeDist} does not exceed 500. This low resolution seems to have a significant effect when modelling mild disruptions with a large pericentre distance. In order to test the hypothesis that a higher resolution would produce a more accurate outcome in this regime, we take a characteristic fragment from the inner annulus in Figure \ref{fig:150AUSizeDist}, i.e. with characteristic values $R=150$ km, $a=0.75$ AU, $q=1R_{\odot}$ and a pure rocky composition. We model this fragment's tidal evolution with full SPH simulations (as done in Paper I), but for only one flyby, using three different resolutions: 0.5K, 5K, and 50K SPH particles. In other words, we test a fragment resolution similar to that of the hybrid model, as well as one-order and two-orders of magnitude greater. The outcomes are shown in Figure \ref{fig:1q0_75a}.

 The relative sizes of the constituent SPH particles are easily noticeable in the first panel (\ref{fig:1q0_75a13000s}), in which the time corresponds to a position outside the Roche limit. Each resolution is also depicted by a different colour: dark-blue for the lowest resolution, blue for the intermediate resolution and light-blue for the highest resolution. Already at this early stage we see a significant difference between the outcomes of each simulation, whereby the fragment distension clearly correlates with the number of its constituent SPH particles. The second panel (\ref{fig:1q0_75a62000s}) captures the stream subsequent to the near-completion of the fragmentation phase, following its gravitational collapse. However, this fragmentation actually occurs \emph{only} in the highest resolution simulation, splitting the original object into three separate sub-fragments (two narrow residual streams are in the process of conjoining with the right and left sub-fragments), while the two other lower resolution simulations show no sign of similar behaviour, and the original object remains whole.

 We conclude that for tidal disruption simulations with a large pericentre distance, near the Roche limit, the resolution must be sufficiently high in order to generate physically plausible results, or else the SPH code requires some other modification. This issue never manifests itself in any of our simulations with pericentre distances of $q=0.5R_{\odot}$ or lower, regardless of resolution. Previous similar studies \citep{VerasEtAl-2014,VerasEtAl-2017} have reported no dependence at all, for any value of $q$, for resolutions beyond merely $\sim$1000 particles, however these authors utilize a very different N-body numerical model. This provides a hint as to the possible cause.

 We have found that a possible solution to this problem is to lower the artificial viscosity parameters in the SPH model. We postulate that under certain conditions, the artificial viscosity, if allowed to operate unchecked, does not operate only when particle convergence occurs, but also acts to slow down neighbouring particles as they overtake because of differential rotation. We have successfully verified that lowering the artificial viscosity parameters by a factor of two fixes the problem in the intermediate resolution simulation. Lowering them by a factor of 4, further fixes the problem even in the lowest resolution simulation. We however remain cautious with regard to the possibility of systematically reducing artificial viscosity in the SPH model (currently we implement fiducial $\alpha$ and $\beta$ values of 1 and 2, respectively, as typically suggested by \cite{SchaferEtAl-2016}), and propose that this merits future investigation of various artificial viscosity treatments.

 How might these issues affect the hybrid model interpretation of results? Conceptually speaking, they are of little consequence. We propose that for the size distribution outcomes of simulations with $q\le 0.5R_{\odot}$ there is no effect at all. When $q$ nears the Roche limit, we must simply postulate a flat distribution outcome, rather than rely on the hybrid model final outcomes, since our ability to increase the resolution indefinitely is unrealistic, and the artificial viscosity treatment requires additional studies. Our investigation in Figure \ref{fig:1q0_75a}, however, proves that the flat distribution outcome is \emph{physically} sound.
\end{appendices}

\bsp	
\label{lastpage}
\end{document}